\documentclass[universe,review,accept,moreauthors,pdftex]{Definitions/mdpi} 
\usepackage{amsmath}
\usepackage{physics}
\usepackage{amsfonts}
\usepackage{amssymb}
\usepackage{mathtools}
\usepackage{graphicx}
\usepackage{slashed}
\usepackage{circuitikz}
\usepackage{comment}
\usepackage{tikz}
\usepackage{caption}
\usepackage[labelformat=simple]{subcaption}  
\DeclareCaptionLabelFormat{subcaptionlabel}{\normalfont(\textbf{#2}\normalfont)}
\usepackage[compat=1.1.0]{tikz-feynman}
\firstpage{1} 
\pubvolume{8}
\issuenum{6}
\articlenumber{299}
\pubyear{2022}
\copyrightyear{2020}
\externaleditor{Academic Editor: Farvah \linebreak Mahmoudi} 
\datereceived{20 March 2022} 
\dateaccepted{17 May 2022} 
\datepublished{25 May 2022} 
\hreflink{https://doi.org/10.3390/\linebreak universe8060299} 
\pdfoutput=1

\Title{Introduction to Charged Lepton Flavor 
	Violation}

\TitleCitation{Introduction to Charged Lepton Flavor 
	 Violation}

\newcommand{\meg}	{\mbox{$\mu^+ \rightarrow e^+ \gamma $ }}
\newcommand{\mueee}		{\mbox{$\mu^\pm \rightarrow e^\pm e^- e^+ $ }}
\newcommand{\mueeep}		{\mbox{$\mu^+ \rightarrow e^+ e^- e^+ $ }}
\newcommand{\muconv}    {\mbox{$\mu^- \mbox{N}\rightarrow e^- \mbox{N} $ }}
\newcommand{\muconvp}    {\mbox{$\mu^- \mbox{N}\rightarrow e^+ \mbox{N$^{\prime}$} $ }}
\newcommand{\taumug}    {\mbox{$\tau^{\pm} \rightarrow \mu^{\pm} \gamma $ }}
\newcommand{\taueg}    {\mbox{$\tau{\pm} \rightarrow e^{\pm} \gamma $ }}
\newcommand{\taulll}    {\mbox{$\tau\rightarrow 3l $ }}
\newcommand{\taulh}    {\mbox{$\tau\rightarrow l+h $ }}

\Author{Marco Ardu $^{1,*,\dagger}$ and Gianantonio Pezzullo $^{2, \dagger}$\orcidA{}}

\AuthorNames{Marco Ardu and Gianantonio Pezzullo}

\AuthorCitation{Ardu, M.; Pezzullo, G.}

\address{%
$^{1}$ \quad Laboratoire Univers et Particules de Montpellier, CNRS, 
 Université Montpellier Place Eugene Bataillon, CEDEX 5, F-34095 Montpellier, France\\
$^{2}$ \quad Physics Department, Yale University, 217 Prospect ST,
 New Haven, CT 06511, USA; g.pezzullo@yale.edu
}

\corres{Correspondence: marco.ardu@umontpellier.fr}
\firstnote{These authors contributed equally to this work.}

\abstract{Neutrino masses are evidence of lepton flavour violation, but no
 violation in the interactions among the charged leptons has been observed yet.
Many models of Physics Beyond the Standard Model (BSM) predict Charged Lepton Flavor Violation (CLFV) in 
a wide spectrum of processes with rates in reach of upcoming experiments.
The experimental searches that provide the current best limits on the 
CLFV searches are reviewed, with a particular emphasis on the muon-based 
experiments that give the most stringent constraints on the BSM parameter space.
The next generation of muon-based experiments (MEG-II, Mu2e, COMET, Mu3e) 
aims to reach improvements by many orders of magnitude with respect to the current 
best limits, thanks to several technological advancements.
We review popular heavy BSM theories, and we present the 
calculations of the predicted CLFV branching ratios, focusing on the more sensitive $\mu\to e$ sector.}


\begin{document}





\section{Introduction}
In the Standard Model (SM) defined with massless left-handed neutrinos, Lepton Flavor (LF) is a conserved quantity. The observation of neutrino oscillations provided clear evidence of non-zero neutrino masses and mixing angles, demonstrating that lepton flavour is not a symmetry of nature. Charged Lepton Flavour Violation (CLFV), defined as a short-range interaction among the charged leptons, is therefore expected to occur but it is yet to be observed. If neutrinos get their masses through renormalizable Yukawa interactions with the Higgs, the predicted rates for CLFV are typically GIM suppressed $G^2_F m^4_\nu$$\sim$10$^{-50}$ and are practically unobservable. A detection of CLFV would thus be a clear signature of new physics that could shed light on the origin of neutrino masses. Additionally, lepton flavour is an accidental symmetry of the SM that is respected by the most general Lagrangian with gauge invariant renormalizable interactions. Thus, independently from neutrino masses, minimal departures from the SM can easily introduce extra sources of lepton flavour violation and lead to sizeable CLFV rates.

For these reasons, experimental searches of CLFV attract great interest and are a valuable tool in identifying viable Beyond Standard Model (BSM) scenarios. CLFV searches can pinpoint theories at energy scales currently not directly accessible by the collider facilities. Null results from the current experiments significantly constrain the parameter space of new physics models, and the improvements in sensitivity by several orders of magnitude, especially in the $\mu\to e$ sector, will further probe BSM physics.

In this article, we present an overview of the theoretical and experimental status of CLFV. 
Excellent complementary reviews on the subject already exist in the literature~\cite{Kuno:1999jp, Bernstein, Signorelli, cei_review, Lindner:2016bgg}. 
In  Section \ref{sec2}, we discuss several SM extensions that could be potentially probed in the upcoming CLFV experimental searches. The focus is on heavy physics models. We briefly review the flavour structure of the SM, and we discuss the LFV phenomenology of models that generate neutrino masses at tree and loop level.  In the subsequent sections, we present the CLFV signature of various BSM scenarios, such as Two Higgs Doublets Model and supersymmetric SM. Finally, LFV is studied in the context of effective field theory where model-independent results are retained. 

In Section \ref{sec:exp}, the state of the art and the upcoming experiments looking for CLFV processes are discussed. A particular emphasis is given to those looking for rare muon CLFV decays. Several facilities around the world (Fermilab, PSI and J-PARC) already started building or commissioning new generation experiments with improved sensitivity on the muon CLFV searches (up to four orders of magnitude). 
This is possible thanks to improvements in the acceleration 
techniques, necessary to deliver beam with unprecedented intensity $\sim$10$^{10}$ \textmu/s, and novel detector technologies.
The same section also provides an overview of the current best limits achieved on the tau CLFV branching ratios set by general-purpose experiments at  $e^+e^-$ and $pp$ colliders. Also on the tau front, an improved sensitivity on several searches is expected thanks to the unprecedented luminosity of the Large Hadron Collider at CERN and the SuperKEKB collider at KEK laboratory.

\section{CLFV in Standard Model Extensions}\label{sec2}
The gauge interactions in the Standard Model (SM) are universal among different families, and the Yukawa interactions of leptons and quarks with the Higgs doublet are the only flavor defining couplings:
\begin{equation}
	-\mathcal{L}_{Yuk}=[Y_e]_{ij}\bar{\ell}_i H e_j+ [Y_u]_{nm}\bar{q}_n \tilde{H} u_m+[Y_d]_{nm}\bar{q}_n H d_m+\text{h.c}
\end{equation}
where $\ell_i=(\nu_{iL}\ e_{iL})^T$, $q_i=(u_{iL}\ d_{iL})^T$ are SU(2)  left-handed doublets, while $u_m,d_m,e_j$ label right-handed up, down quark and lepton, respectively. $H$ is the Higgs doublet $H=(H^{+}\ H_0)^T$, and $\tilde{H}$ is defined as $\tilde{H}\equiv \epsilon H^\dagger$, having introduced the SU(2) antisymmetric tensor $\epsilon_{12}=-\epsilon_{21}=1$. We define the electric charge as $Q=T_3+Y$, where $Y$ is the hypercharge and $T_3$ is the diagonal generator of SU(2). 
The labels $i,j,n,m$ denote the generation indices, and the Yukawa matrices $Y_e,Y_u,Y_d$ are $3\times3$ complex matrices
 in flavor space. Any complex matrix can be diagonalized with a bi-unitary transformation
\begin{equation}
	Y_u=V_{q_u} \hat{Y}_u V^{\dagger}_u\qquad Y_d=V_{q_d} \hat{Y}_d V^{\dagger}_d\qquad Y_e=V_l \hat{Y}_e V^{\dagger}_e 
\end{equation}
where $\hat{Y}_f$ are diagonal matrices with non-negative entries. The quark up and down Yukawa matrices cannot be simultaneously diagonalized in the unbroken theory, given that $V_{q_d}\neq V_{q_u}$. In a basis where the up Yukawa matrix is diagonal, the down Yukawa can be cast in the following form:
\begin{equation}
	Y_d= V^\dagger_{q_u} V_{q_d}D_d\equiv V D_d
\end{equation}
where $V$ is the Cabibbo-Kobayashi-Masukawa (CKM) \textls[-15]{unitary matrix, which parameterizes flavour violation in the quark sector. When the Electroweak gauge symmetry is spontaneously broken by the Vacuum Expectation Value (VEV) of the Higgs doublet $\expval{H}=\begin{pmatrix} 0 & v\end{pmatrix}^T$, with $v\simeq 174$ GeV, the SM fermions acquire masses through the Yukawa~interactions}
\begin{equation}
	m_f=vY_f\qquad \text{where 
}\ f=e,u,d
\end{equation}
so that the mass eigenstate basis $d'_L$ for left-handed down quark is related to the gauge interaction basis by a CKM rotation
\begin{equation}
	d_{iL}=V_{ij} d'_{jL}
\end{equation}

Only the charged currents are affected by a unitary rotation on $d$ quarks, and the flavour-changing interactions with the $W$ bosons are governed by the CKM matrix elements
\begin{equation}
	\mathcal{L}_W= -\frac{g}{\sqrt{2}}W^+_\alpha V_{ij}\bar{u}_{iL}\gamma^\alpha d'_{jL}+\text{h.c}
\end{equation}
where $g=e (\sin\theta_W)^{-1}$ is the $SU(2)$ gauge coupling. In the lepton sector, the Yukawa couplings $Y_e$ are the only basis-choosing interaction, and lepton flavour $U(1)_{L_e}\otimes U(1)_{L_\mu}\otimes U(1)_{L_\tau}$ is conserved because mass and gauge eigenstate basis coincide
. (The symmetry is anomalous and is not exact at the non-perturbative level. However, $B/3-L_{e,\mu,\tau}$ is anomaly free, hence non-baryonic processes like $\mu\to e \gamma$ are strictly forbidden in the SM). This holds as long as neutrinos are massless and do not provide additional flavour defining~interactions. 

Since the so-called ``solar neutrino problem'' of the 1960s, a deficit \cite{DeficitSolarNeutrino1,DeficitSolarNeutrino2, DeficitSolarNeutrino3, DeficitSolarNeutrino4, DeficitSolarNeutrino5, DeficitSolarNeutrino6} in the number of electron neutrinos compared to the prediction of the standard solar model \cite{NeutrinoSolarCalc1, NeutrinoSolarCalc2, NeutrinoSolarCalc3},  neutrino oscillations have been confirmed by many observations \cite{Oscillation1,Oscillation2, Oscillation3, Oscillation4, Oscillation5} that firmly established non-zero and non-degenerate neutrino masses. Weak eigenstates are superpositions of mass eigenstates
\begin{equation}
	\nu_{iL}=U_{i1}\nu_{1L}+U_{i2}\nu_{2L}+U_{i3}\nu_{3L}\qquad \text{with}\ i=e,\mu,\tau
\end{equation}
where $U$ is known as Pontecorvo-Maki-Nakagawa-Sakata (PMNS) \cite{PMNS1,PMNS2} matrix, which is the lepton analogue of CKM. It is a unitary matrix, and it is parametrized by nine real parameters, three angles and six phases. Not all phases are physical, and some can be absorbed in field phase redefinition. The canonical form of PMNS is the following:
\begin{equation}
	U=\begin{pmatrix}
		c_{12}c_{13} & s_{12}c_{13} & s_{13}e^{-i\delta}\\
		-s_{12}c_{23}-c_{12}s_{23}s_{13}e^{i\delta} & c_{12}c_{23}-s_{12}s_{23}s_{13}e^{i\delta} & s_{23}c_{13}\\
		s_{12}s_{23}-c_{12}c_{23}s_{13}e^{i\delta} & -c_{12}s_{23}-s_{12}c_{23}s_{13}e^{i\delta} & c_{23}c_{13}
	  \end{pmatrix}\times P
\end{equation}
where we have defined $s_{ij}=\sin\theta_{ij},c_{ij}=\cos\theta_{ij}$ and the matrix $P$ is the identity if neutrinos have Dirac masses, while it contains two extra phases for Majorana neutrinos $P=\text{diag}\begin{pmatrix}1 & e^{i\alpha_{12}} & e^{i\alpha_{31}}\end{pmatrix}$. This is because, with self-conjugate left-handed neutrinos, there are less relative field re-definitions that can absorb the matrix phases. Majorana phases are difficult to observe because they contribute as extra sources of CP violation in processes that depend linearly on the masses, hence not in oscillations where the dependence is quadratic. A recent global fit \cite{deSalas:2020pgw} on neutrino oscillation data gives the following parameters for mass-squared differences and mixing angles:
\begin{align}
	m^2_{2}-m^2_{1}=\left[6.94-8.14\right]\times 10^{-5}\ e\text{V}^2\qquad \abs{m^2_{3}-m^2_{1}}=\left[2.47-2.63\right]\times 10^{-3}\ e\text{V}^2 \nonumber\\
	\sin^2\theta_{12}=\left[2.71-3.69\right]\times 10^{-1}\qquad \sin^2\theta_{23}=\left[4.34-6.1\right]\times 10^{-1} \\ 
	\sin^2\theta_{13}=\left[2.000 - 2.405\right]\times 10^{-2}\qquad \delta=\left[0.71 - 1.99\right]\times \pi \nonumber
\label{eq:NeutrinoFIT}
\end{align}
where the 3$\sigma$ range are displayed and normal ordering $m_1<m_2<m_3$ is assumed. The sign of $\Delta m^2_{21}$ is determined by solar neutrino observation because relevant matter effects depend on it. On the other hand, atmospheric neutrino data measure only the absolute value of mass squared difference $\Delta m^2_{31}$, and they can be consistent with inverted ordering $m_3<m_1<m_2$, which leads to marginally different best-fit values for the PMNS parameters. 
\subsection{CLFV in Models That Generate Neutrino Mass at Tree Level}
Assuming the presence of three right-handed neutrinos $\nu_{Ri}$ which are singlets of the SM gauge group, gauge invariance allows for Yukawa couplings between the lepton and the Higgs doublet that generate Dirac masses for neutrinos when electroweak symmetry is spontaneously broken
\begin{equation}
	-\mathcal{L}_\nu=[Y_{\nu}]_{ij}\bar{\ell}_i \tilde{H} \nu_{Rj}+\text{h.c}.
\end{equation}

To obtain neutrino masses that are compatible with cosmological constraints  \linebreak $m_\nu\lesssim$ 0.1 eV~\cite{Nuetrinomasscosmology}, the neutrino Yukawa couplings must be $Y_\nu\lesssim \order{10^{-12}}$. Although small Yukawas are technically natural 
 (with technical naturalness we refer to naturalness as defined by t'Hooft. Small values for a parameter $c$ are defined as technically natural when a symmetry is restored in taking the limit $c\to 0$), Dirac masses require a strong hierarchy between the charged and neutral lepton Yukawa sector.

Analogously to CKM, the PMNS matrix is the result of the misalignment between charged lepton and neutrino mass basis, as the neutrino and charged lepton Yukawas cannot be simultaneously diagonalized respecting the electroweak gauge symmetry. \textls[-25]{Flavor violation is hence parameterized by the presence of the PMNS matrix in the charged lepton~currents }
\begin{equation}
		\mathcal{L}_W= -\frac{g}{\sqrt{2}}W^-_\alpha \sum_{\substack{i=e,\mu,\tau \\ j=1,2,3}}U_{ij}\bar{e}_{iL}\gamma^\alpha \nu_{jL}+\text{h.c}\label{eq:LFVdirac}
\end{equation}

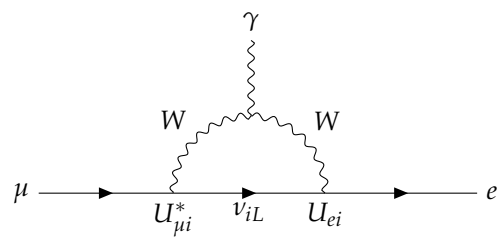
\begin{figure}[H]
	\begin{tikzpicture}
		\begin{feynman}[node distance=2 cm]
			\vertex (mu) {\(\mu\)};
			\vertex [right=of mu] (a) [label=-90:\(U^*_{\mu i}\)];
			\vertex [above right=1.45 cm of a](midph);
			\vertex [right= of a] (b) [label=-90:\(U_{e i}\)];
			\vertex [right=of b] (e) {\(e\)};
			\vertex [above=1 cm of midph] (ph) {\(\gamma\)};
			
			\diagram* [small]{
				(mu) -- [fermion] (a) -- [fermion, edge label'=\(\nu_{iL}\)] (b) -- [fermion] (e),
				(a) -- [boson, quarter left, edge label=\(W\)] (midph) -- [boson, quarter left, edge label=\(W\)] (b),
				(midph) -- [boson] (ph),
			};
		\end{feynman}
	\end{tikzpicture}
	\caption{$\mu\to e \gamma$ mediated by massive neutrinos $\nu_{iL}$. }
	\label{fig:mutoeneutrinos}
\end{figure}

Charged lepton flavour violation is mediated by neutrino mixing. In Figure~\ref{fig:mutoeneutrinos}, we show a representative diagram for $\mu\to e \gamma$ decay, illustrating that this is a consequence of the flavour-changing interactions in Equation~(\ref{eq:LFVdirac}). The amplitude of this process can be generically cast in the following form 
 (see, for instance, Chapter 6, Section 6.2 of \cite{Peskin:1995ev})):
\begin{align}
	\mathcal{M}(\mu\to e \gamma)&=\bar{u}_e(p_e)(m_\mu(A_RP_R+A_LP_L)i\sigma_{\alpha \beta}q^\beta + (B_RP_R+B_LP_L)q_\alpha+ \\
	&+(C_RP_R+C_LP_L)\gamma_\alpha)u_\mu(p_e+q)\epsilon^{*\alpha}(q)\nonumber\\
	&=\mathcal{M}_\alpha \epsilon^{*\alpha}(q)\nonumber
\end{align}
where $P_{R,L}=(1\pm \gamma_5)/2$ are the right-handed and left-handed chiral projectors, while $A,B,C$ are complex numbers.
As a consequence of QED gauge invariance, the amplitude satisfies the Ward identity $q^\alpha \mathcal{M}_\alpha=0$. On-shell spinors obey the equation of motion $(\slashed{p}-m)u(p)=0$, and the Ward identity requires
\begin{equation}
	m_\mu(C_R P_R+C_L P_L)-m_e(C_RP_L+C_LP_R)+q^2(B_RP_R+B_LP_L)=0
\end{equation}
which, for on-shell photons $q^2=0$, has the unique solution $C_R=C_L=0$. The only relevant piece is a dipole transition 
\begin{equation}
	\mathcal{M}(\mu\to e \gamma)=\bar{u}_e(p_e)\left[i\sigma_{\alpha \beta}q^\beta m_\mu(A_RP_R+A_LP_L)\right]u_\mu(p_e+q)\epsilon^{*\alpha}(q) \label{eq:amplitudemutoegamma}
\end{equation}
which is chirality-flipping and, thus, proportional to the muon mass (neglecting the electron mass). Equation (\ref{eq:amplitudemutoegamma}) yields the following decay rate \cite{Kuno:1999jp}:
\begin{equation}
	\Gamma(\mu\to e \gamma)=\frac{m_\mu^5}{16\pi}(\abs{A_L}^2+\abs{A_R}^2)
\end{equation}
\vspace{-10 pt}
In the diagram of Figure \ref{fig:mutoeneutrinos}, the outgoing electrons are left-handed and only $A_R$ is non-zero. The amplitude is proportional to the internal neutrino propagator, which can be expanded for small neutrino masses as
\begin{equation}
	\sum_i \frac{U^*_{ei} U_{\mu i}}{(k^2-m^2_i)}=\sum_i \frac{U^*_{ei} U_{\mu i}}{k^2}+\sum_i \frac{U^*_{ei} U_{\mu i}}{k^2}\left(\frac{m_i^2}{k^2}\right)+\order{\frac{m_i^4}{k^4}}\label{eq:GIM}.
\end{equation}

We see that the leading term vanishes due to PMNS unitarity, and the amplitude is GIM suppressed by the square of neutrino masses. Indeed, the process is analogous to a flavour-changing neutral current in the quark sector, which features a similar GIM suppression by CKM unitarity \cite{GIM}. 
The calculation is done in \cite{Cheng:1984vwu} in the $R_\xi$ gauge, where additional diagrams replacing $W$ with the charged Goldstones must be included. All diagrams are finite and in the unitary gauge limit $\xi\to \infty$; only the diagram of Figure \ref{fig:mutoeneutrinos} is non-zero. Dividing by the rate $\Gamma(\mu\to e \nu\bar{\nu})=G_F^2 m^5_\mu/192\pi^3$ of the dominant LF conserving three-body decay, the resulting branching ratios for $\mu\to e \gamma$ is \cite{Mutoegamma1, Mutoegamma2,Mutoegamma3,Mutoegamma4, Lee:1977tib}
\begin{equation}
	Br(\mu\to e\gamma)=\frac{3\alpha_{e}}{32\pi}\abs{\sum_i U^*_{ei} U_{\mu i}\frac{m_{i}^2}{M_W^2}}^2.
\end{equation}

Rewriting the sum as
\begin{equation}
	\sum_i U^*_{ei} U_{\mu i}\frac{m_{ i}^2}{M_W^2}=U^*_{e2} U_{\mu 2}\frac{\Delta m_{21}^2}{M_W^2}+U^*_{e3} U_{\mu 3}\frac{\Delta m_{31}^2}{M_W^2}
\end{equation}
\textls[-25]{and substituting the best-fit values of the mass differences and mixing parameters in \mbox{Equation~(9)}, the predicted branching ratios for the LFV $\mu\to e \gamma$ is $Br(\mu\to e \gamma)= 10^{-54}-10^{-55}$,} which lie beyond any foreseeable experimental reach. In models with Dirac neutrino masses, rates of other LFV processes are similarly GIM suppressed and, thus, too small to be observable. 

The right-handed neutrinos are \textit{sterile}, i.e., neutral under the SM gauge group, so SM gauge invariance allows us to add a lepton number violating Majorana mass term
\begin{equation}
	-\mathcal{L}_\nu=[Y_{\nu}]_{ij}\bar{\ell}_i \tilde{H} \nu_{Rj}+\frac{1}{2}[M_R]_{ij} \overline{\nu^c}_{iR} \nu_{jR}+\text{h.c}.
\end{equation}
having defined $\nu^c=C\bar{\nu}^T$, where $C$ is the Dirac charge conjugation matrix \cite{ChargeConjugation}. Majorana mass matrices are symmetric because fermion fields are anti-commuting and the charge conjugation matrix $C$ is antisymmetric.  Upon electroweak symmetry breaking, the mass Lagrangian can be cast in the following form (suppressing generation indices) 
\begin{equation}
	-\mathcal{L}_\nu=\frac{1}{2}\overline{N^c} M_{N} N+\text{h.c}\qquad \text{where}\ M_N=\begin{pmatrix}
		0 & M_D\\
		M^T_D & M_R
	\end{pmatrix}
\end{equation}
where $N=\begin{pmatrix}
	\nu^c_L &
	\nu_R
\end{pmatrix}^T$ and $M_D=vY_\nu$. If we assume that the Majorana masses $M_R$ are much larger than the Dirac masses (symbolically $M_R\gg M_D$), the matrix can be put in block diagonal form that disentangles light left-handed neutrinos and heavy right-handed~ones~\cite{Grimus:2000vj}
\begin{equation}
W^T M_N W=\begin{pmatrix}
	M_{\nu} & 0\\
	0 & M_{{heavy}}
\end{pmatrix},\qquad \begin{pmatrix}
	\nu^c_L \\
	\nu_R
\end{pmatrix}=W\begin{pmatrix}
\nu^c_{light} \\
\nu_{heavy}
\end{pmatrix}
\end{equation}
where $W$ is a unitary matrix. 
At leading $M_DM^{-1}_R$ order, the mass matrices are
\begin{equation}
	M_{{heavy}}=M_R\qquad M_{\nu}=-M_DM^{-1}_RM^T_D
	\label{eq:seesawformula}
\end{equation}

Assuming $\sim\order{1}$ Yukawas, light neutrinos masses can be explained by Majorana masses close to the grand unification scale $M_R\sim$10$^{15}$ GeV. This is the celebrated seesaw mechanism \cite{Seesaw}, specifically known as type I when the SM is extended with singlet right-handed fermions.

A unitary $U^*$ diagonalizes the symmetric Majorana matrix $M_\nu$ with a congruence transformation $
	U^{*T} M_\nu U^*=\hat{M}_\nu=\text{diag}\begin{pmatrix}
		m_1 & m_2 & m_3
	\end{pmatrix}$,
but $U$ is not the matrix that appears in the charged currents. Defining $U\equiv U\otimes 1_{heavy}$ as acting on the light neutrinos subspace, gauge interaction and mass basis are related by
\begin{equation}
	\begin{pmatrix}
		\nu^c_L \\
		\nu_R
	\end{pmatrix}=WU^*\begin{pmatrix}
		\nu^c_{light} \\
		\nu_{heavy}
	\end{pmatrix},
\end{equation}
where the matrix $W$ can be expanded at second order as \cite{Grimus:2000vj}
\begin{equation}
	W=\begin{pmatrix}
		1-\frac{1}{2}B_1B^\dagger_1 & B_1\\
		-B^\dagger_1 & 1-\frac{1}{2}B^\dagger_1B_1 
	\end{pmatrix}\qquad \text{with}\ B_1=(M_R^{-1}M_D^T)^\dagger=v(M_R^{-1}Y_\nu^T)^\dagger .
\end{equation}

The left-handed weak eigenstates are related to the light mass eigenstates through a non-unitary matrix $U'$
\begin{equation}
	\nu_L=U'\nu_{light}=\left(1-\frac{1}{2}(B_1B^\dagger_1)^* \right)U\nu_{light}=\left(1-\frac{v^2}{2}Y_\nu\frac{1}{M^\dagger_R M_R}Y^\dagger_\nu \right)U\nu_{light} \label{eq:nonunitary}
\end{equation} 

Lacking unitarity, the GIM suppression no longer operates substituting the $U'$ matrix in Equation~(\ref{eq:GIM}), and the rate of $\mu \to e \gamma$ becomes \cite{ Cheng:1984vwu, Antusch:2006vwa}
\begin{equation}
	\frac{\Gamma(\mu\to e\gamma)}{\Gamma(\mu\to e\nu\bar{\nu})}=\frac{3\alpha_{e}}{32\pi}\frac{\abs{\sum_i U'^*_{ei} U'_{\mu i}F(x_i)}^2}{(U'U'^\dagger )_{ee}(U'U'^\dagger )_{\mu\mu}}
\end{equation}
where $x_i=m^2_i/M_W^2$ and $F(x_i)$ is a loop function that can be expanded for $x_i\ll 1$ as  $F(x_i)\simeq 10/3-x_i$. CLFV processes can thus constrain departures from the unitarity of the PMNS matrix \cite{XingZhangUnitarity}. Substituting the typical value $y_\nu^2\simeq m_\nu M_R/v^2$, GIM suppression is replaced by the ratio $m_\nu/M_N$, which for GUT scale sterile neutrinos predict rates that are nonetheless well below future experimental sensitivity. Seesaw models can predict sizeable CLFV rates if the Majorana right-handed masses are closer to the electroweak scale. In the non-supersymmetric seesaw, this is also desirable to avoid large correction to the Higgs mass \cite{Vissani:1997ys}. However, in a generic setup with T$e$V scale $M_R$ and unsuppressed CLFV rates, fine-tuned cancellations are required in Equation~(\ref{eq:seesawformula}) to explain neutrino masses. Fine-tuning is, of course, avoided if a symmetry principle forces the neutrino mass to be small despite having large Yukawa couplings. Observe that neutrino masses are a lepton number violating effect $\sim Y_\nu M^{-1}_R Y^T_\nu$, while the non-unitary matrix that governs CLFV rates is lepton number conserving $\sim Y_\nu M^{-2}_R Y_\nu^\dagger$. It is possible to suppress neutrino masses by invoking a small breaking of lepton number conservation while keeping the masses of the sterile neutrinos sufficiently close to the electroweak scale and with no need for small Yukawa couplings. This is, for example, achieved in the inverse seesaw \cite{Inverseseesaw1,Inverseesaw2} and by considering a quasi-degenerate pair of sterile neutrinos \cite{Pseudodirac1, Pseudodirac2, PseudoDirac3}.

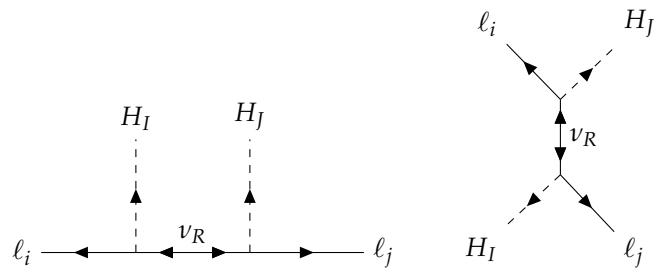
\begin{figure}[H]
	
	\begin{tikzpicture}
		\begin{feynman}
			\vertex (nu) {\(\ell_{i}\)};
			\vertex [right=of nu] (a) ;
			\vertex [above=of a] (H1) {\(H_I\)};
			\vertex [right= of a] (b);
			\vertex [above=of b] (H2) {\(H_J\)};
			\vertex [right=of b] (nu2) {\(\ell_{j}\)};
			
			\diagram*[small] {
				(nu) -- [anti fermion] (a) -- [anti majorana, edge label=\(\nu_R\)] (b) -- [fermion] (nu2),
				(a) -- [charged scalar] (H1),
				(b) -- [charged scalar] (H2),
			};
		\end{feynman}
	\end{tikzpicture}\qquad
	\begin{tikzpicture}
		\begin{feynman}[node distance=1cm]
			\vertex (a);
			\vertex [above left=of a](nu) {\(\ell_{i}\)};
			\vertex [above right=of a] (H1) {\(H_J\)};
			\vertex [below= of a] (b);
			\vertex [below left=of b] (H2) {\(H_I\)};
			\vertex [below right=of b] (nu2) {\(\ell_{j}\)};
			
			\diagram*[small] {
				(nu) -- [anti fermion] (a) -- [anti majorana, edge label=\(\nu_R\)] (b) -- [fermion] (nu2),
				(a) -- [charged scalar] (H1),
				(b) -- [charged scalar] (H2),
			};
		\end{feynman}
	\end{tikzpicture}
	\caption{Matching contributions to the Weinberg operator in type I seesaw.}
	\label{fig:weinberg}
\end{figure}

\begin{figure}[H]

	\begin{tikzpicture}
		\begin{feynman}
			\vertex (nu) {\(\ell_{i}\)};
			\vertex [right=of nu] (a) ;
			\vertex [above=of a] (H1) {\(H_I\)};
			\vertex [right= of a] (b);
			\vertex [above=of b] (H2) {\(H_J\)};
			\vertex [right=of b] (nu2) {\(\ell_{j}\)};
			
			\diagram* [small]{
				(nu) -- [anti fermion] (a) -- [anti majorana, edge label=\(\Sigma\)] (b) -- [fermion] (nu2),
				(a) -- [charged scalar] (H1),
				(b) -- [charged scalar] (H2),
			};
		\end{feynman}
	\end{tikzpicture}\qquad
	\begin{tikzpicture}
		\begin{feynman}[node distance=1cm]
			\vertex (a);
			\vertex [above left=of a](nu) {\(H_{I}\)};
			\vertex [above right=of a] (H1) {\(H_J\)};
			\vertex [below= of a] (b);
			\vertex [below left=of b] (H2) {\(\ell_i\)};
			\vertex [below right=of b] (nu2) {\(\ell_{j}\)};
			
			\diagram* [small]{
				(nu) -- [anti charged scalar] (a) -- [anti charged scalar, edge label=\(\Delta\)] (b) -- [fermion] (nu2),
				(a) -- [charged scalar] (H1),
				(b) -- [fermion] (H2),
			};
		\end{feynman}
	\end{tikzpicture}
	\caption{Seesaw Majorana neutrino masses generated by integrating out a heavy scalar triplet $\Delta$ (type II) or a heavy fermion triplet $\Sigma$ (type III). }
	\label{fig:seesaws}
\end{figure}
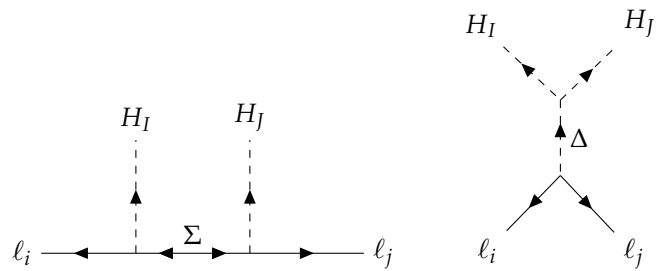

The seesaw formula can be understood as the result of integrating out the heavy  neutrinos. The relevant $s$ and $t$ channel diagrams are shown in Figure \ref{fig:weinberg} and match onto the dimension five Weinberg operator
\begin{equation}
	\mathcal{L}_{d=5}=\frac{1}{2}C^5_{ij}(\bar{\ell}_i\tilde{H})(\ell^c_j \tilde{H})+\text{h.c}
\end{equation}
with the coefficient $C^5=Y_\nu M^{-1}_R Y^T_\nu$. When the Higgs doublet gets a VEV, neutrinos acquire Majorana masses through the Weinberg operator $M_\nu\equiv -v^2C^5=-M_DM^{-1}_RM^T_D$, that agrees with Equation~(\ref{eq:seesawformula}). Moreover, the following dimension six operator is generated
\begin{equation}
	\mathcal{L}_{d=6}=C^6_{ij}(\bar{\ell}_i \tilde{H})i\slashed{\partial}(\tilde{H}^\dagger \ell_j),
\end{equation}
with a coefficient $C_6=Y_\nu M_R^{-1\dagger}M_R^{-1} Y^\dagger_{\nu}$, which corrects the light neutrinos kinetic terms. The redefinition needed to canonically normalize the fields introduces a non-unitary matrix in the charged currents \cite{Abada:2007ux}
\begin{equation}
\nu_L\to (\delta_{ij}+v^2C_{ij}^6)^{-1/2}\nu_L\ \rightarrow\ 
		\mathcal{L}_W= -\frac{g}{\sqrt{2}}W^-_\alpha \sum_{\substack{i=e,\mu,\tau \\ j=1,2,3}}\left( \delta_{ik}-\frac{v^2}{2}C^6_{ik}\right)U_{kj}\bar{e}_{iL}\gamma^\alpha \nu_{jL}+\text{h.c}\label{eq:LFVseesaw}
\end{equation} 
which again agrees with Equation~(\ref{eq:nonunitary}). The advantage of an effective field theory description is that different seesaw scenarios can be described at low energy in a common framework. In Figure \ref{fig:seesaws}, we show how extending the Standard Model with particles transforming in different representations of the SM gauge group can generate Majorana neutrino masses via the Weinberg operator. Recent effective field theory analysis of type I and type II seesaw models include the complete one-loop matching onto effective operators \cite{Zhang:2021jdf,Li:2022ipc}, which are useful resources to study the low-energy CLFV signatures.  For more complete reviews on the CLFV phenomenology of seesaw models, we refer the reader to \cite{Abada:2007ux, Hambye:2013jsa}.

\vspace{-10pt} 
\subsection{CLFV in Models That Generate Neutrino Masses at Loop Level}
New physics not too far from the electroweak scale can account for small neutrino masses if they are generated radiatively via loop diagrams. As a specific example, the so-called scotogenic model \cite{Ma:2006km} adds an additional scalar doublet $\eta$ (with hypercharge $Y_\eta=1/2$) to the Standard Model, together with three generations of sterile neutrinos $N$. The new particles are assumed to be odd under a discrete $Z_2$ symmetry, which forbids Yukawa couplings with the SM Higgs between lepton doublets and the sterile neutrinos, as well as constraining the possible interactions in the scalar potential. The $Z_2$ is also responsible for keeping stable the lightest new particle, which, if neutral, provides a potential dark matter candidate.
Omitting the kinetic terms, the scotogenic Lagrangian reads 
 (it is always possible to diagonalize the symmetric Majorana mass matrix of the sterile neutrinos with no loss of generality)
\begin{equation}
	\mathcal{L}_{sc}=\mathcal{L}_{\rm SM}+\left([Y_\eta]_{ij}\bar{\ell}_i \tilde{\eta}  N_{j}-\frac{M_{Ni}}{2}\overline{N^c}_{i} N_{i}+\text{h.c}\right)-V(H,\eta)
\end{equation}
where the scalar potential is
\begin{align}
	V(H,\eta)&=m_h^2 H^\dagger H+m_\eta^2 \eta^\dagger \eta+\frac{\lambda_1}{2}(H^\dagger H)^2+ \frac{\lambda_2}{2}(\eta^\dagger \eta)^2+\lambda_3 (H^\dagger H)(\eta^\dagger \eta) \nonumber \\
	 &+\lambda_4 (H^\dagger\eta)(\eta^\dagger H)+\frac{\lambda_5}{2}\left[(H^\dagger\eta)^2+(\eta^\dagger H)^2\right].
\end{align}

To preserve the $Z_2$ symmetry when the electroweak symmetry is spontaneously broken, the potential parameters must be such that $\eta$ field does not acquire a VEV. We also assume that all parameters in the potential are real and CP is conserved. With this assumption, the real and imaginary parts of the uncharged component $\eta_0=(\eta_R+i\eta_I)/\sqrt{2}$ do not mix. The mass splitting between the two neutral scalars is proportional to $\lambda_5 v^2$, consequently $\eta_{R,I}$ are approximately degenerate in the limit $\lambda_5\ll 1$. Note that lepton number is conserved if $\lambda_5$ is zero, so small values are technically natural.  

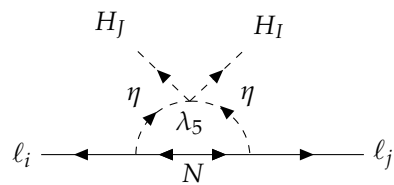
\begin{figure}[H]
	\begin{tikzpicture}
		\begin{feynman}
			\vertex (mu) {\(\ell_i\)};
			\vertex [right=of mu] (a) ;
			\vertex [above right=1 cm of a] (midph) [label=-90:\(\lambda_5\)];
			\vertex [right= of a] (b);
			\vertex [right=of b] (e) {\(\ell_j\)};
			\vertex [above right=1cm of midph] (H1) {\(H_I\)};
			\vertex [above left=1cm of midph] (H2) {\(H_J\)};
			
			\diagram* [small]{
				(mu) -- [anti fermion] (a) -- [anti majorana, edge label'=\(N\)] (b) -- [fermion] (e),
				(a) -- [ charged scalar, quarter left, edge label=\(\eta\)] (midph) -- [anti charged scalar, quarter left, edge label=\(\eta\)] (b),
				(midph) -- [charged scalar] (H1),
				(midph) -- [charged scalar] (H2),
			};
		\end{feynman}
	\end{tikzpicture}
	\caption{Radiative neutrino mass in the scotogenic model. The loop and $\lambda_5$ suppression allows for T$e$V$-$scale new physics and small neutrino masses.   }
	\label{fig:scotogenicmass}
\end{figure}

The $Z_2$ symmetry prevents the appearance of tree-level Majorana masses for the left-handed neutrinos, but it can generate them at the one-loop level 
 through the $\lambda_5$ mixing of $\eta$ with the SM Higgs doublet, as shown in the diagram of Figure \ref{fig:scotogenicmass}. The resulting neutrino mass matrix is calculable, and, for $\lambda_5\ll 1$ ($m_0=m_{\eta_R}\sim m_{\eta_I}$), can be approximated as \cite{Ma:2006km}   
\begin{equation}
    [M_\nu]_{ij}\simeq\frac{2\lambda_5 Y_{\eta ik}Y_{\eta jk}v^2}{16\pi^2 M_{N_i}}\left[\frac{M^2_{N_k}}{m^2_0-M^2_{N_k}}+\frac{M^4_{N_k}}{(m^2_0-M^2_{N_k})^2}\log(\frac{M^2_{N_k}}{m^2_0})\right].	
\end{equation}

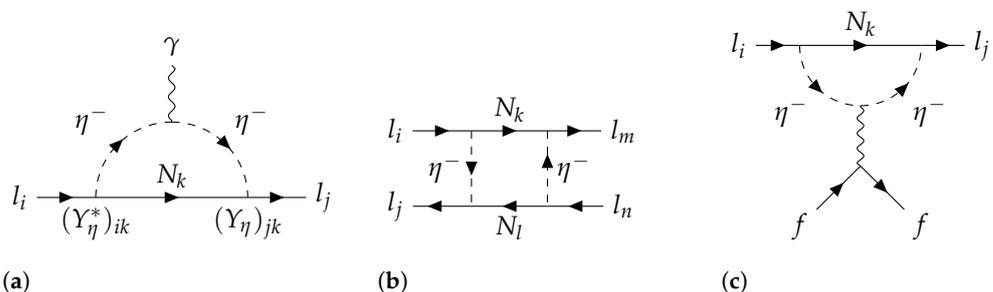
\begin{figure}[H]
\begin{subfigure}{.35\textwidth}
	\begin{tikzpicture}
	\begin{feynman}[small]
		\vertex (mu) at (-1,0) {\(l_i\)};
		\vertex (e) at (3,0) {\(l_j\)};
		\vertex (a)  at (0,0) [label=-90:\((Y^*_{\eta})_{ik} \)];
		\vertex (b) at (2,0) [label=-90:\((Y_{\eta})_{jk} \)];
		\vertex (boson1) at (1, 1);
		\vertex (boson2) at (1, 2) {\(\gamma\)};
		\vertex (c) at (1,0) ;
		\diagram* [inline=(a.base)]{
			(mu) -- [fermion] (a),
			(b) -- [fermion] (e),
			(a) -- [charged scalar, quarter left, edge label=\(\eta^-\)] (boson1) --  (boson1) -- [charged scalar, quarter left, edge label=\(\eta^-\)] (b),
			(boson1) -- [photon] (boson2),
			(a) -- [fermion, edge label=\(N_k\)] (b)
		};
	\end{feynman}
\end{tikzpicture}
\caption{}\label{subfig:scotdipole}
\end{subfigure}
\begin{subfigure}{.32\textwidth}
\begin{tikzpicture}
	\begin{feynman}[small]
		\vertex (l1) at (-1,1) {\(l_i\)};
		\vertex (l2) at (-1,0) {\(l_j\)};
		\vertex (l3) at (2,1) {\(l_m\)};
		\vertex (l4) at (2,0) {\(l_n\)};
		\vertex (e1) at (0,1);
		\vertex (e2) at (0,0);
		\vertex (e3) at (1,1);
		\vertex (e4) at (1,0);
		\diagram* [inline=(a.base)]{
			(l1) -- [fermion] (e1),
			(e1) -- [fermion, edge label=\(N_k\)] (e3),
			(e3) -- [fermion] (l3),
			(l2) -- [anti fermion] (e2),
			(l4) -- [fermion] (e4),
			(e4) -- [fermion, edge label=\(N_l\)] (e2),
			(e1) -- [charged scalar, edge label'=\(\eta^-\)] (e2),
			(e3) -- [anti charged scalar, edge label=\(\eta^-\)] (e4),
		};
	\end{feynman}
\end{tikzpicture}
\caption{}\label{subfig:scotbox}
\end{subfigure}
\begin{subfigure}{.3\textwidth}
\begin{tikzpicture}[scale=0.8]
	\begin{feynman}[small]
		\vertex (mu) at (-1,0) {\(l_i\)};
		\vertex (e) at (3,0) {\(l_j\)};
		\vertex (a)  at (0,0);
		\vertex (b) at (2,0);
		\vertex (f1) at (0,-3) {\(f\)};
		\vertex (f2) at (2,-3) {\(f\)};
		\vertex (boson1) at (1, -1);
		\vertex (boson2) at (1, -2);
		\vertex (c) at (1,0) ;
		\diagram* [inline=(a.base)]{
			(mu) -- [fermion] (a),
			(b) -- [fermion] (e),
			(a) -- [charged scalar, quarter right, edge label'=\(\eta^-\)] (boson1) --  (boson1) -- [charged scalar, quarter right, edge label'=\(\eta^-\)] (b),
			(boson1) -- [photon] (boson2),
			(a) -- [fermion, edge label=\(N_k\)] (b),
			(f1) -- [fermion] (boson2) -- [fermion] (f2),
		};
	\end{feynman}
\end{tikzpicture}
\caption{}\label{subfig:scotpenguin}
\end{subfigure}
\caption{CLFV processes in the scotogenic model.  ({\bf a}) Diagrams contributing to the $l_i\to l_j\gamma$ rate.
({\bf b}) Box diagrams contributing to the $l_i\to l_m l_n l_j$ rate.
({\bf c}) Penguin diagrams contributing to the $l_i\to l_j l_k l_k$ rate and $\mu\to e$ conversion rate ($f$ can be a quark or a lepton). }\label{fig:LFVscotogenic}
\end{figure}

With respect to the traditional seesaw scenario, the extra suppression $\sim\lambda_5/(16\pi^2)$ can predict small vales of $m_\nu$ with T$e$V scale sterile neutrinos and unsuppressed Yukawa couplings. The CLFV signature of the scotogenic model have been studied  with particular attention to $l_i\to l_j \gamma$ processes \cite{Kubo:2006yx,AristizabalSierra:2008cnr,Suematsu:2009ww}, while the phenomenology of $l_i\to l_jl_kl_m$ and $\mu \to e$ conversion into nuclei has also been discussed \cite{Toma:2013zsa}. In Figure \ref{fig:LFVscotogenic}, we show a selection of diagrams giving contributions to CLFV processes at the one-loop level.  

\vspace{-12pt} 

Part of the parameter space of the scotogenic model is excluded by the current experimental LFV searches, while the viable region can give branching ratios within upcoming experimental sensitivities and will be probed in the near future. It is often the case that $l_i\to l_j \gamma$ is the most constraining LFV channel because the dipole (Figure \ref{fig:LFVscotogenic}a) contribution to the photon penguin (Figure \ref{fig:LFVscotogenic}c) can dominate the amplitude of $l_i\to l_j \bar{f}f$, leading to the following relation \cite{Arganda:2005ji}
\begin{equation}
    Br(l_i\to 3l_j )\sim \frac{\alpha_{\rm em}}{3\pi}\left(2\log(\frac{m_{l_i}}{m_{l_j}})-\frac{11}{4}\right) Br(l_i\to l_j \gamma)
\end{equation}

\begin{figure}[H]
\begin{subfigure}{0.46\textwidth}
    \includegraphics[scale=0.21]{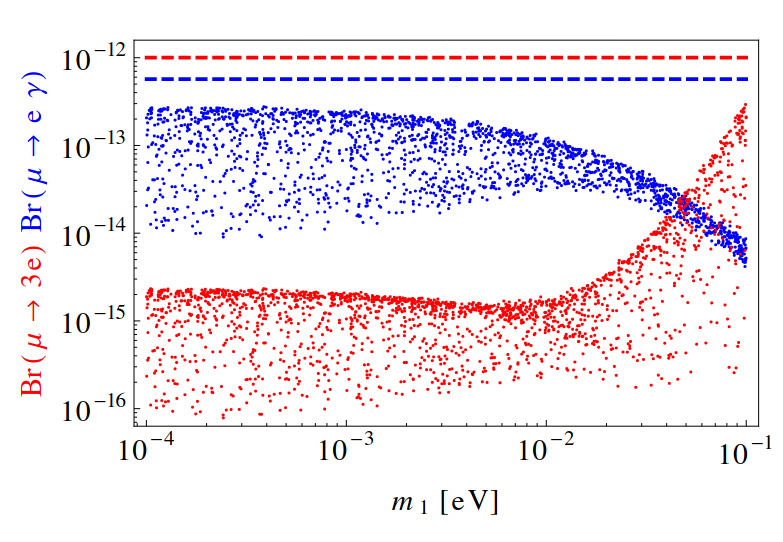}
\end{subfigure}
\begin{subfigure}{0.3\textwidth}
    \includegraphics[scale=0.21]{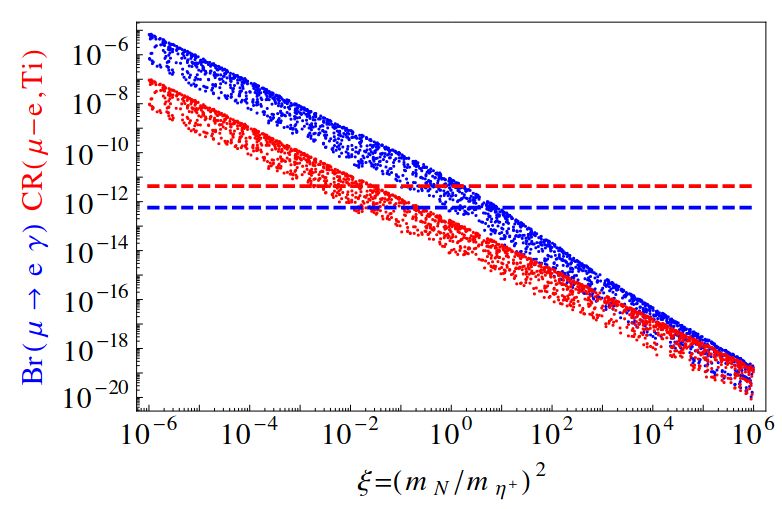}
\end{subfigure}
    \caption{The plots show some CLFV branching fractions in the scotogenic model: in the left plot the prediction for $Br(\mu\to e \gamma)$ and $Br(\mu\to 3e)$ for degenerate sterile neutrino masses $m_N=4$ TeV and $m_{\eta^{+}}=1$ TeV, varying the mass of the lightest neutrino (normal ordering); in the right figure $Br(\mu\to e \gamma)$ and $Br(\mu \to e)$ conversion as a function of $(m_N/m_{\eta^+})^2$. The dashed lines correspond to the current experimental upper limit. Yukawas $Y_\eta$  compatible with neutrino parameters are randomly generated (Figure from 
 \cite{Toma:2013zsa}).}\label{fig:scotogenic}
\end{figure}

However, the box contribution (Figure \ref{fig:LFVscotogenic}b) can be larger than the photon penguin diagram for mass of the lightest neutrino close to the cosmological upper limit 0.1 eV (Figure~\ref{fig:scotogenic}) so that upcoming $\mu\to 3e$ searches can constrain the model orthogonally to the MEG bound on $\mu\to e\gamma$. 
In addition, the penguin diagram of Figure \ref{fig:LFVscotogenic}c mediates LFV interactions with quarks, contributing to the rate of $\mu\to e$ conversion in nuclei (we briefly review the $\mu\to e$ conversion rate calculation in Appendix \ref{appendix:mutoeconv}).
When the dipole dominates the penguin amplitude, $\mu\to e$ conversion experimental reach is not competitive with $\mu\to e \gamma$ searches 
 (although with the future branching ratios sensitivity $Br(\mu A\to e A) \sim 10^{-16}$, $\mu\to e$ conversion might be able to probe smaller dipole coefficients than MEG II with $Br(\mu\to e\gamma)\sim 6 \times 10^{-14}$), given that
\begin{equation}
    \frac{Br(\mu\to e\gamma)}{Br(\mu N(A,Z)\to e N(A,Z))}\sim f(A,Z)\times 10^{2} \label{eq:MEGvsCONVscot}
\end{equation}
where $f(A,Z)$ is a nucleus dependent factor that is $\sim \mathcal{O}(1)$ for the targets used in experiments \cite{Kuno:1999jp}. As shown in the right plot of Figure \ref{fig:scotogenic}, the scaling of Equation~(\ref{eq:MEGvsCONVscot}) is satisfied for small $m_N/m_\eta$ ratios, while the non-dipole penguin amplitude can come to dominate for larger $m_N/m_\eta$ values. The upcoming $\mu\to e$ conversion searches will be a valuable probe for this region of parameter space \cite{Vicente:2014wga}.

Another popular model that can generate neutrino masses at loop level is the Zee-Babu Model \cite{Zee1986QuantumNO, BABU1988132}, where the SM is extended with two SU(2) singlet and charged scalars $k^+, k^{++}$ and allow for the Lagrangian terms
\begin{equation}
    \mathcal{L}_{ZB}\supset f^+_{ij}\overline{\ell^c}_{iI}\epsilon_{IJ}\ell_{jJ}k^++f^{++}_{ij}\overline{e^c}_{i}e_{j}k^{++},
\end{equation}
where $\epsilon_{IJ}$ is the anti-symmetric $SU(2)$ tensor. Lepton number is not conserved and neutrino masses are generated at the two-loop level,
 while the interactions also violate lepton flavor. The CLFV phenomenology of the Zee-Babu model has been studied in \cite{Herrero-Garcia:2014hfa,Nebot:2007bc,AristizabalSierra:2006gb}. 
For other models that generate neutrino masses at loop level and their CLFV signatures, we refer the reader to \cite{Cai:2017jrq}. 

\subsection{Two Higgs Doublet Model}
Neutrino masses are just one among pieces of evidence and hints of new physics. Dark matter, the baryon asymmetry, $B$ anomalies, the hierarchy and the strong CP problem are some of the observations and theoretical conundrums that call for SM extensions. CLFV is a feature of numerous models that address the above problems and can be employed to constrain or, if CLFV is ever observed, determine the region of parameter space where Beyond Standard Model theories must sit. 

One simple extension of the Standard Model features an additional scalar doublet $H_2$, which is commonly known as the Two Higgs Doublet Model (2HDM) (for a review see \cite{Branco:2011iw}). A second Higgs is strongly motivated by supersymmetry, where one Higgs cannot give masses to all fermions and the second Higgsino, the superpartner of the second doublet, is necessary to cancel the gauge anomalies. Although supersymmetry imposes precise relations among the Higgs masses and couplings, supersymmetry breaking terms lead to modifications so that, at low energy, it is suitable to describe the two Higgs with generic couplings. A general 2HDM (type III) predicts LFV couplings that must be sufficiently suppressed to satisfy the current experimental constraint. Often, additional symmetries are assumed to avoid the appearance of flavour-changing neutral current at tree level. In Type I 2HDM, the SM fermions only couple to one Higgs, while, in the type II, the up quarks couple to a different Higgs with respect to leptons and down quarks, which is the case for supersymmetric SM.

The 2HDM scalar Lagrangian is the following
\begin{align}
	-\mathcal{L}_{\rm 2HDM}&=[Y_e]_{ij}\bar{\ell}_i H_1 e_j+ [Y_u]_{nm}\bar{q}_n \tilde{H}_1 u_m+[Y_d]_{nm}\bar{q}_n H_1 d_m+\text{h.c}\nonumber \\
	&+[K_e]_{ij}\bar{\ell}_i H_2 e_j+ [K_u]_{nm}\bar{q}_n \tilde{H}_2 u_m+[K_d]_{nm}\bar{q}_n H_2 d_m+\text{h.c} \\
	&+V(H_1, H_2) \nonumber
\end{align}
where the potential reads
\begin{align}
	V(H_1,H_2)&=m^2_{11}H^\dagger_{1}H_1+m^2_{22}H^\dagger_{2}H_2-m^2_{21}(H^\dagger_1H_2+\text{h.c})+\frac{\lambda_1}{2}(H^\dagger_1 H_1)^2+\frac{\lambda_2}{2}(H^\dagger_2 H_2)^2\nonumber \\
	&+\lambda_3 (H^\dagger_1 H_1)(H^\dagger_2 H_2)+\lambda_4(H^\dagger_1 H_2)(H^\dagger_2 H_1)\nonumber \\
	&+\left(\frac{\lambda_5}{2}(H^\dagger_1 H_2)^2
	+\lambda_6(H^\dagger_1 H_1)(H_1H^\dagger_2)+\lambda_7(H^\dagger_2 H_2)(H_1H^\dagger_2)+\text{h.c}\right).
\end{align}

In a region of the potential parameters, the Higgs acquire a VEV that spontaneously breaks the electroweak gauge symmetry, and it is always possible to rotate in a basis where only one has a non-zero expectation value $\expval{H_1}=\begin{pmatrix} 0& v \end{pmatrix}^T$, $\expval{H_2}=0$. The doublets are written~as
\begin{equation}
	H_1=\begin{pmatrix}
		G^{+} \\
		v+\frac{1}{\sqrt{2}}(\rho_1+iG^0)
	\end{pmatrix}\qquad H_2=\begin{pmatrix}
	\phi^{+} \\
	\frac{1}{\sqrt{2}}(\rho_2+iA)
\end{pmatrix}.
\end{equation} 

Once the Goldstones $G$ are eaten by the gauge bosons, the scalar spectrum contains one $\phi^{+}$ complex scalar, two CP even neutral scalar $\rho_{1,2}$ and one CP odd scalar $A$. If the potential parameters are real, only the two CP even scalars mix, and we identify two mass eigenstates  $h, H$ \cite{Sacha2HDM1,Sacha2HDM2}
\begin{align}
	h=\sin(\beta - \alpha)\rho_1+\cos(\beta - \alpha)\rho_2\equiv s_{\beta\alpha}\rho_1+c_{\beta\alpha}\rho_2  \\
	H=\cos(\beta - \alpha)\rho_1-\sin(\beta - \alpha)\rho_2\equiv c_{\beta\alpha}\rho_1-s_{\beta\alpha}\rho_2
\end{align}
where $\beta-\alpha$ is the angle that diagonalizes the neutral scalar mass matrix, and $h$, $H$ have masses $m_h<m_H$, respectively. We identify the lighter scalar with the 125$-$GeV Higgs boson. In a basis where the two doublets $\Phi_i$ both have VEVs $v_i$, a rotation with angle $\alpha$ diagonalizes the neutral scalar mass matrix, while the angle $\beta$ such that $\tan \beta\equiv  v_1/v_2$ allows us to rotate into the $H_i$ basis.  In the Type III 2HDM, there is no unambiguous way to identify $\beta$ and $\alpha$ because the two doublets are not distinguishable. On the other hand, $\beta-\alpha$ is calculable in terms of the potential parameters \cite{Sacha2HDM2} \begin{equation}
	\cos(\beta-\alpha)\sin(\beta-\alpha)=-\frac{2\lambda_6v^2}{(m^2_H-m_h^2)}\label{eq:Higgsbasisangle}
\end{equation}
In the fermion mass basis, the Yukawa matrices $Y_f$ are diagonalized, while the $K_f$ couplings with $H_2$ are, in general, non-diagonal. The Yukawa interactions between the fermions and the uncharged scalar sector read 
\begin{align}
	-\mathcal{L}_{Y}=\frac{h}{\sqrt{2}}\bar{f}_{iL}\left(\frac{[m_f]_{i}\delta_{ij}}{v}s_{\beta\alpha}+[K_f]_{ij}c_{\beta\alpha}\right)f_{jR}+\text{h.c}\nonumber\\
	\frac{H}{\sqrt{2}}\bar{f}_{iL}\left(\frac{[m_f]_{i}\delta_{ij}}{v}c_{\beta\alpha}-[K_f]_{ij}s_{\beta\alpha}\right)f_{jR}+\text{h.c}\nonumber\\
    \sum_{f=d,e}i\frac{A}{\sqrt{2}}\bar{f}_{iL}[K_f]_{ij}f_{jR}-i\frac{A}{\sqrt{2}}\bar{u}_{iL}[K_u]_{ij}u_{jR}+\text{h.c}
\end{align}

The LHC measure a $h\to \tau^+\tau^-,\mu^+\mu^-$ rate compatible with the Standard Model prediction \cite{ATLAS:2018ynr,CMS:2017zyp}, requiring $s_{\beta\alpha}\sim 1$. Substituting this approximation in Equation~(\ref{eq:Higgsbasisangle}) gives
\begin{equation}
	c_{\beta\alpha}\simeq -2\lambda_6v^2/(m^2_H-m_h^2)\ll 1\ \to \ c_{\beta\alpha}\simeq\frac{-2\lambda_6 v^2}{m_H^2}.
\end{equation} 

In the decoupling limit $\lambda_iv^2\ll m^2_{22}$ \cite{Gunion:2002zf} that we have assumed to justify $m^2_H\gg m^2_h$, the mass splitting $m^2_{H}-m^2_A\sim \lambda_5 v^2$ is small, and in the following, we consider $M^2\sim m_H^2\sim m^2_A$. The off-diagonal interaction $K_{e}c_{\beta\alpha}$ can mediate LFV Higgs boson decay with a rate \cite{2HDMHiggsDecay, 2HDMHiggsLFVdecay}
\begin{equation}
    \Gamma(h\to l_i l_j)= \frac{\abs{K_{e}}_{ij}^2+\abs{K_{e}}_{ji}^2}{16\pi}c_{\beta\alpha}^2 m_h\qquad \text{where
}\ l_il_j=l^+_il^-_j+l^-_il^+_j
\end{equation}

Non-observation of LFV decay modes of the Higgs boson at LHC set the upper limits on the branching fractions reported in Table \ref{tab:higgsdecayLFV} and directly constrain the size of flavour violating coupling.
\begin{table}[H]
  \caption{Lepton flavour violating decay of the SM Higgs boson with the current experimental bounds set by ATLAS and CMS.}
   \begin{tabularx}{\textwidth}{m{5cm}<{\centering}m{7cm}<{\centering}}
\toprule
        \textbf{Process} & \textbf{Bound} \\
        \midrule
        $h\to\mu e$  & $6.1\times10^{-5}$ \cite{ATLAS:2019xlq}\\
        \midrule
        $h\to \tau \mu$ & $1.5\times 10^{-3}$ \cite{cms_htaul}\\
        \midrule
        $h\to \tau e$ & $2.2\times 10^{-3}$ \cite{cms_htaul}\\
\bottomrule
    \end{tabularx}
  
    \label{tab:higgsdecayLFV}
\end{table}

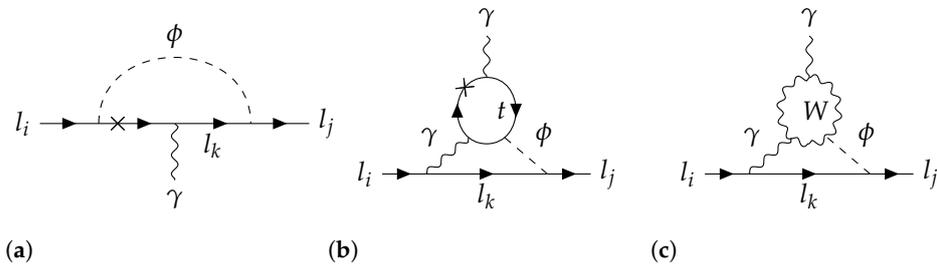
\begin{figure}[H]
\begin{subfigure}{.3\textwidth}
	\centering
\begin{tikzpicture}
	\begin{feynman}[small]
		\vertex (mu) at (-1,0) {\(l_i\)};
		\vertex (e) at (3,0) {\(l_j\)};
		\vertex (a)  at (0,0) ;
		\vertex (b) at (2,0) ;
		\vertex (boson1) at (1.5, 1);
		\vertex (boson2) at (2, 2);
		\vertex (c) at (1,0) ;
		\vertex (H3) at (1,-1) {\(\gamma\)};
		\diagram* [inline=(a.base)]{
			(H3) -- [photon] (c),
			(mu) -- [fermion] (a),
			(b) -- [fermion] (e),
			(a) -- [scalar, half left, edge label=\(\phi\)] (b) -- [anti fermion, edge label=\(l_k\)] (c) -- [anti fermion, insertion=0.75] (a),
		};
	\end{feynman}
\end{tikzpicture}
\caption{}\label{subfig:oneloopHIGGS}
\end{subfigure}
\begin{subfigure}{.3\textwidth}
	\centering
\begin{tikzpicture}[scale=0.8]
	\begin{feynman}[small]
		\vertex (mu) at (-1,0) {\(l_i\)};
		\vertex (e) at (3,0) {\(l_j\)};
		\vertex (a)  at (0,0) ;
		\vertex (b) at (2,0) ;
		\vertex (c) at (1,0) ;
		\vertex (loop1) at (1,0.5);
		\vertex (loop2) at (1,1.6);
		\vertex (photon) at (0.7,0.6);
		\vertex (scalar) at (1.3,0.6);
		\vertex (photon2) at (1, 2.6) {\(\gamma\)};
		\diagram* [inline=(a.base)]{
			(mu) -- [fermion] (a),
			(b) -- [fermion] (e),
			(a) -- [fermion, edge label'=\(l_k\)] (b),
			(loop1) -- [fermion, half left, insertion=0.75] (loop2) -- [fermion, half left, edge label'=\(t\)] (loop1),
			(a) -- [photon, edge label=\(\gamma\)] (photon),
			(b) -- [scalar, edge label'=\(\phi\)] (scalar),
			(loop2) -- [photon] (photon2),
		};
	\end{feynman}
\end{tikzpicture}
\caption{}\label{subfig:twoloopHIGGS1}
\end{subfigure}
\begin{subfigure}{.3\textwidth}
	\centering
\begin{tikzpicture}[scale=0.8]
	\begin{feynman}[small]
		\vertex (mu) at (-1,0) {\(l_i\)};
		\vertex (e) at (3,0) {\(l_j\)};
		\vertex (a)  at (0,0) ;
		\vertex (b) at (2,0) ;
		\vertex (c) at (1,0) ;
		\vertex (loop1) at (1,0.5);
		\vertex (loop2) at (1,1.6);
		\vertex (photon) at (0.7,0.6);
		\vertex (scalar) at (1.3,0.6);
		\vertex (photon2) at (1, 2.6) {\(\gamma\)};
		\diagram* [inline=(a.base)]{
			(mu) -- [fermion] (a),
			(b) -- [fermion] (e),
			(a) -- [fermion, edge label'=\(l_k\)] (b),
			(loop1) -- [photon, half left] (loop2) -- [photon, half left, edge label'=\(W\)] (loop1),
			(a) -- [photon, edge label=\(\gamma\)] (photon),
			(b) -- [scalar, edge label'=\(\phi\)] (scalar),
			(loop2) -- [photon] (photon2),
		};
	\end{feynman}
\end{tikzpicture}
\caption{}\label{subfig:twoloopHIGGS2}
\end{subfigure}
\caption{Diagrams for $l_i\to l_j\gamma$ in the 2HDM, where $\phi=h,H,A$. Two-loop Bar-Zee diagrams with a $Z$ exchange also exist. 
({\bf a}) One loop contribution to the $l_i\to l_j \gamma$ rate in the 2HDM with LFV Yukawa couplings.
({\bf b}) Two loop Barr-Zee diagram with a top loop contributing to $l_i\to l_j \gamma$ in the 2HDM with LFV Yukawa couplings.
({\bf c}) Two loop Barr-Zee diagram with a W loop contributing to $l_i\to l_j \gamma$ in the 2HDM with LFV Yukawa couplings. }\label{fig:2DHMliljgamma}
\end{figure}

\begin{figure}[H]
\begin{subfigure}{0.43\textwidth}
    \includegraphics[scale=0.25]{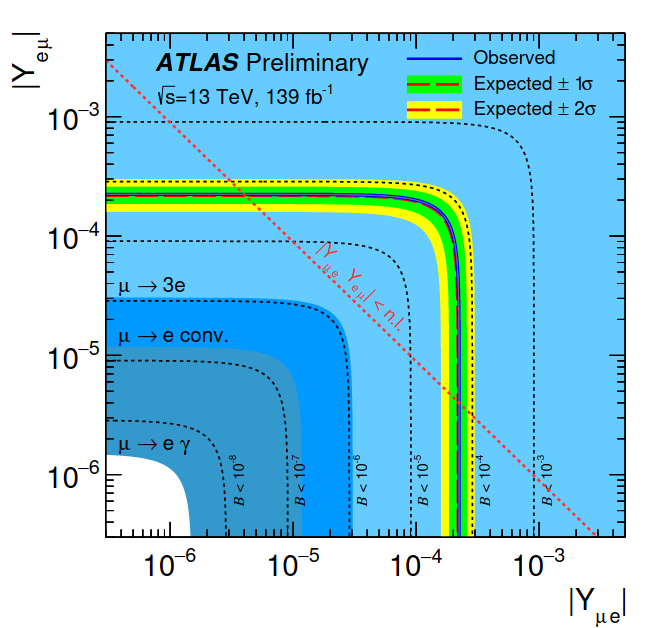}
\end{subfigure}
\begin{subfigure}{0.36\textwidth}
    \includegraphics[scale=0.33]{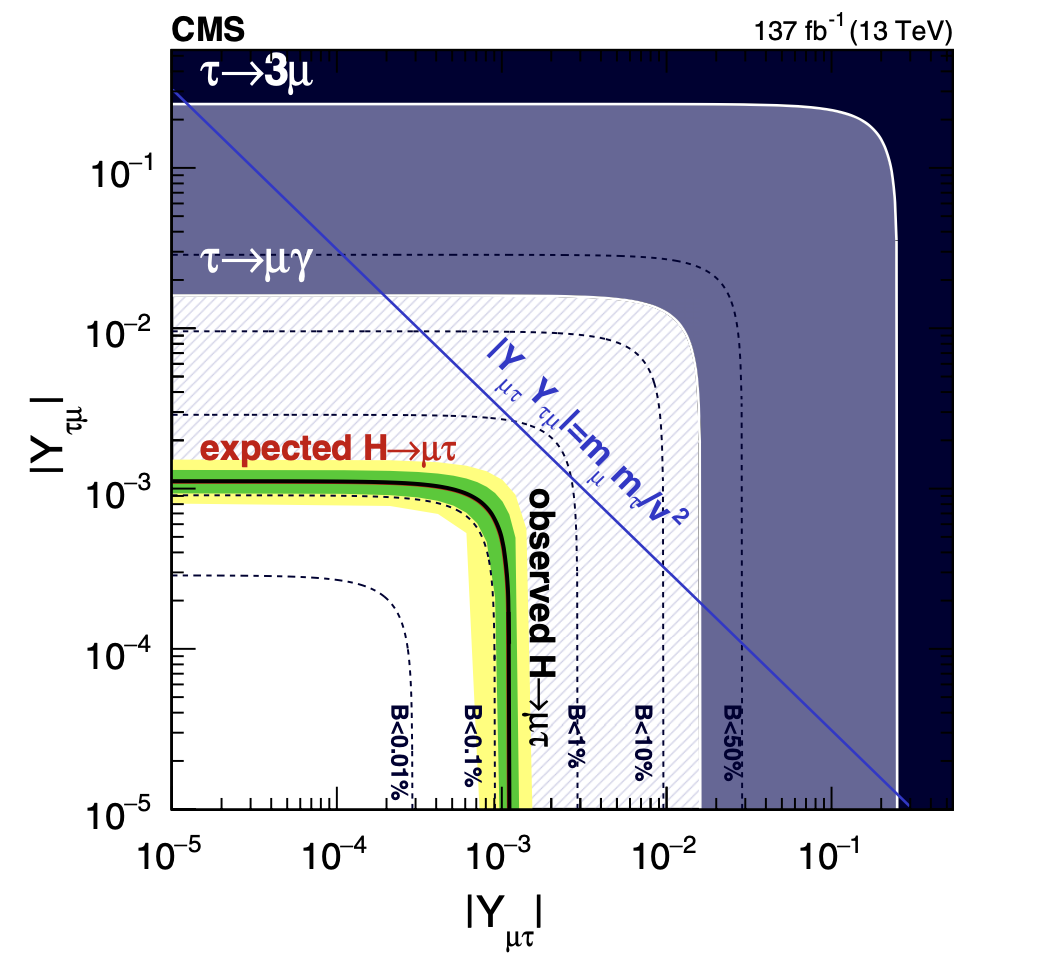}
\end{subfigure}
    \caption{Left 
 figure from 
 \cite{ATLAS:2019xlq}: constraint on LFV Yukawa couplings $Y_{\mu e}, Y_{e \mu}$ from the limits on $Br(h\to e\mu)$ (observed limit corresponds to the solid blue line, while the expected one is the dashed red line). Shaded regions show the sensitivity of $\mu\to 3e, \mu\to e \gamma$ and $\mu\to e$ conversion on the LFV Yukawas, from the calculations of  \cite{2HDMZupan}.  Right figure from \cite{cms_htaul}: same plot for the $\tau\leftrightarrow \mu$ sector. The diagonal line shows the natural limit  $\abs{Y_{ij}Y_{ji}}<2m_im_j/v^2$ \cite{AnsatzChengSher}.  }\label{fig:HiggsLFV}
\end{figure}

Off-diagonal Yukawas are also indirectly bounded by other LFV processes, as they can mediate $l_i\to l_j\gamma$ through the loop diagrams shown in Figure \ref{fig:2DHMliljgamma}. The two-loop diagrams of Figure \ref{fig:2DHMliljgamma}b,c are relevant and can be numerically larger than one-loop contributions \cite{Bjorken:1977vt} because, in the former, the Higgs line is attached to a heavy particle running in the loop and Yukawa suppression is avoided. In the $\mu\to e$ sector, $\mu\to e \gamma$ is the most sensitive process to LFV Yukawas, which has been extensively studied in the context of 2HDM \cite{Sacha2HDM2, 2HDM2, 2HDM3, 2HDM4, 2HDM5}. For the $\tau\leftrightarrow l$ sector, the bound on the radiative decay $Br(\tau\to l\gamma)<\text{few}\times10^{-8}\to 10^{-9}$ is less stringent and the Higgs LFV decays are sensitive to smaller off-diagonal Yukawas. In a simplified scenario where only the SM Higgs is present, the author of \cite{2HDMZupan} computed several processes in terms of generic LFV couplings $hY_{ij}\bar{e}_{Li}e_{Rj}$, and in Figure \ref{fig:HiggsLFV}, we show the current bounds (sensitivity) set by LFV observables. In 2HDM, there are also contributions from the heavy scalars, which are parametrically of similar size with respect to light Higgs LFV; while they do not suffer from the small mixing angle $c_{\beta\alpha}\sim v^2/M^2$, the propagator yields a similar suppression $\sim 1/M^2$. 

\vspace{-10pt} 

Contribution to $l_i\to l_{j}l_kl_k$ and  $l_i \to l_j qq$ also appear at tree level in the 2HDM, but they are suppressed by the small diagonal Yukawas. The same processes receive relevant contributions by attaching a $l_jl_j$ ($q q$) current to the photon of the diagrams in Figure \ref{fig:2DHMliljgamma}.

\subsection{CLFV in Supersymmetry}
Supersymmetry (SUSY) is a space-time symmetry that extends Poincare invariance by adding fermionic generators that satisfy the anti-commutation relations of the supersymmetry algebra \cite{SUSY1, SUSY2, SUSY3}. SUSY is the largest space-time symmetry that the $S-$matrix can have given a set of physical assumptions such as unitarity, locality and causality \cite{ColemanMandela, HaagLopuSUSYalgebra}. Since fermionic operators $Q$ are added to the algebra, irreducible SUSY representation (supermultiplets) contain particles of different spin that are related by the action of $Q$ on one-particle states. In the case of $N=1$ SUSY, i.e., with only one pair of conjugate Weyl spinor generators, the spectrum contains boson-fermion degenerate pairs. In the Minimal Supersymmetric SM (MSSM), for every quark $q$ and lepton $\ell$, there is a corresponding complex scalar in the same gauge representation, commonly known as the squark $\tilde{q}$ and slepton $\tilde{\ell}$. Similarly, the gauginos $\tilde{B},\tilde{W}^I,\tilde{G}^a$ are the fermion superpartner of the gauge bosons, which transforms in the adjoint of the gauge group, while the higgsinos $\tilde{H}_u,\tilde{H}_d$ are the spin-1/2 particles that belong to the supermultiplets of the Higgs doublets. As already discussed in the previous section, a supersymmetric version of the SM requires at least two Higgs doublets.

One of the most attractive features of SUSY is that it provides a solution to the hierarchy problem. The Higgs in the SM is a fundamental scalar, and its mass is quadratically sensitive to any new physics that couples to it. If we assume that a new bosonic or fermionic heavy state with masses $\sim\Lambda$ couples to the Higgs, its mass will get corrections that are quadratic in $\Lambda$. Therefore, the Higgs tree-level mass must cancel almost precisely with the UV contribution, leaving a small remnant mass at the electroweak scale $m_h\sim$125 GeV. If $\Lambda$ is taken near the Planck scale $M_{Pl}\sim10^{19}$ GeV, the fine-tuning required for this accidental cancellation is extreme. If nature is supersymmetric, then the corrections to the Higgs mass cancel precisely among the contributions of superpartners. That is because SUSY predicts the same couplings for all particles in the supermultiplet, and closed fermion loops get minus signs from Grassmann traces. The absence of quadratic divergences can also be understood by observing that fermion-boson pairs are degenerate, and fermion masses are protected by the chiral symmetry and are only logarithmically sensitive to the UV scale.

Degenerate partners of the known SM particles with opposite statistics have never been observed; therefore, SUSY, if realized at all, must be broken at some scale $m_{0}$. To avoid the reappearance of the hierarchy problem, $m_0$ should not be too far from the Higgs mass and explicit SUSY breaking terms must be soft, i.e., have to contain only terms with strictly positive mass dimension. The solution to the hierarchy problem is also preserved if SUSY is spontaneously broken by the expectation value $\sim m_0$ of some scalar field. In models of spontaneous breaking, soft breaking terms appear in the low-energy non-supersymmetric description. Null results from the LHC rule out SUSY breaking scales below few $\times$ TeV \cite{SUSYLHC1, SUSYLHC2, SUSYLHC3}, although the bounds on superpartners' masses are not completely model-independent.

In the MSSM, the SUSY breaking sector can be a source of lepton flavour violation. The soft breaking terms contain masses for the sleptons and trilinear couplings with the~Higgs
\begin{equation}
	-\mathcal{L}_{\text{soft}}\supset [\tilde{m}^2_R]_{ij} \tilde{e}^\dagger_i \tilde{e}_j+[\tilde{m}^2_L]_{ij} \tilde{\ell}^\dagger_i \tilde{\ell}_j+m_0[A]_{ij}\tilde{\ell}^\dagger_i H_d \tilde{e}_j 
\end{equation}
that introduce LFV if the off-diagonal entries are non-zero in the lepton mass eigenstate basis. For $\sim100$ GeV$-$T$e$V soft terms, the current bounds on LFV, and more generally on flavour-changing neutral current in the SM, call for a suppression mechanism of sfermions mass mixing. This is known as the SUSY flavour problem. The spontaneous symmetry breaking of SUSY cannot be triggered by the scalar fields in the MSSM supermultiplets, as this would lead to an unacceptable spectrum. Supersymmetry breaking may occur in a hidden sector that has a very small coupling with the MSSM particle. This is known as the mediation paradigm. If SUSY breaking is communicated via supergravity couplings of the hidden sector to matter, it results in universal and flavour-conserving soft terms at the Planck scale \cite{PhysRevLett.49.970, Barbieri1982GaugeMW}. Nonetheless, this does not strictly forbid LFV, since mass mixing can still be radiatively generated. In minimal SU(5) Grand Unified Theory (GUT), the matter content of SM is reproduced by three generations of a fermion field in the anti-fundamental $\bar{\textbf{5}}$ 
 of SU(5) ($\bar{F}$), which contains the lepton doublet and the right-handed down-type quark,  and one that fills the $\textbf{10}$ representation of SU(5) ($T$), which contains the quark doublet, the right-handed up-type quark and the right-handed charged lepton. They are coupled in the Yukawa sector to two Higgs scalar fields transforming in the $\bar{\textbf{5}}$ and $\textbf{5}$ representation. 
 (For the following, we adopt the common convention in SUSY of dropping Dirac notation, and we use chiral Weyl fermions. The product of left-handed Weyl fermion contracted with the anti-symmetric tensor  $\psi_\alpha \epsilon^{\alpha\beta}\chi_\beta\equiv \psi \chi=\chi\psi$ is Lorentz-invariant). 
\begin{equation}
   -\mathcal{L}_{\rm SU(5), Yuk}=[Y_u]_{ij}T_{i}HT_{j}+[Y_d]_{ij}\bar{F}_{i}\bar{H}T_{j}+\rm h.c
\end{equation}
 
In SUSY GUT, the above equation corresponds to the superpotential $W$, where $T, \bar{F}, H, \bar{H}$ are  the superfields that contain the SM particles and the superpartners.
Assuming gravity-mediated SUSY breaking, the soft terms at $M_{Pl}$ are flavor blind and characterised by a common mass scale $m_0$
\begin{equation}
   -\mathcal{L}_{\rm SU(5), soft}=m^2_0(\tilde{T}^\dagger_{i}\tilde{T}_{i}+\tilde{\bar{F}}^\dagger_{i}\tilde{\bar{F}}_{i})+m_0a_0([Y_u]_{ij}\tilde{T}_{i}H\tilde{T}_{j}+[Y_d]_{ij}\tilde{\bar{F}}_{i}\bar{H}\tilde{T}_{j}+\rm h.c)
\end{equation}

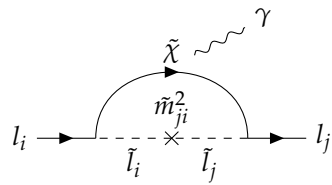
\begin{figure}[H]
\begin{tikzpicture}
	\begin{feynman}[small]
		\vertex (mu) at (-1,0) {\(l_i\)};
		\vertex (e) at (3,0) {\(l_j\)};
		\vertex (a)  at (0,0) ;
		\vertex (b) at (2,0) ;
		\vertex (boson1) at (1.5, 1);
		\vertex (boson2) at (2, 2);
		\vertex (c) at (1.3,1.1) ;
		\vertex (c1) at (1,0) [label=90:\(\tilde{m}^2_{ji} \)];
		\vertex (H3) at (2.2,1.6) {\(\gamma\)};
		\diagram* [inline=(a.base)]{
			(H3) -- [photon] (c),
			(mu) -- [fermion] (a),
			(b) -- [fermion] (e),
			(a) -- [fermion, half left, edge label=\(\tilde{\chi}\)] (b) -- [scalar, edge label=\(\tilde{l}_j\)] (c1) -- [scalar, insertion=0, edge label=\(\tilde{l}_i\)] (a),
		};
	\end{feynman}
\end{tikzpicture}
\caption{$l_i\to l_j \gamma$ in SUSY through sleptons
	 mass mixing. $\tilde{\chi}$ correspond to charginos and neutralinos (mass eigenstates of electroweak gauginos and higgsinos). }\label{fig:SusyCLFV}
\end{figure}

The top Yukawa is large and loop correction to third generation masses can be sizeable. In a basis where the up Yukawa matrix is diagonalized and neglecting first and second generation couplings, the leading-log correction in the renormalization of the $\tilde{T}_3$ mass is~\cite{Hisano:1996qq, Kuno:1999jp}
\begin{equation}
	\Delta \tilde{m}_{T,33}\simeq -\frac{3}{8\pi^2}\abs{Y_u}^2_{33}m_0^2(3+\abs{a_0}^2)\log(\frac{M_{\rm Pl}}{M_{\rm GUT}})
\end{equation}
where $M_{\rm GUT}\sim 10^{16}$ GeV is the GUT scale. The $\tilde{T}$ fields contain the right-handed charged sleptons 
 that have a diagonal but non-universal mass matrix. In the mass eigenstate basis for the charged leptons, the right-handed slepton mass matrix acquires non-diagonal~entries
\begin{equation}
		[\Delta\tilde{m}_{R}]_{ij}\simeq -\frac{3}{8\pi^2}[V^*_e]_{i3}[V_e]_{j3}\abs{Y_u}^2_{33}m_0^2(3+\abs{a_0}^2)\log(\frac{M_{\rm Pl}}{M_{\rm GUT}})\qquad \text{with}\ Y_e=V_{\ell}\hat{Y}_eV^\dagger_e 
\end{equation}
where $\hat{Y}_e$ is the diagonal lepton Yukawa. In SU(5) GUT, the down and lepton Yukawa are unified $Y_e=Y^T_d$, and $V_e$ correspond to the transpose CKM matrix. 
In the diagrams of Figure~\ref{fig:SusyCLFV}, we show how slepton mass mixing can mediate $l_i\to l_j \gamma$ at loop level, which in most SUSY setups is the largest LFV signal. Box diagrams exist for $l_i\to l_j l_k l_k$ and $l_i\to l_j qq$, but the processes are often dominated by the penguin diagrams, where a flavour diagonal current is attached to an off-shell photon in the diagrams of Figure \ref{fig:SusyCLFV}. The rate of $\mu\to e \gamma$ in minimal SU(5) GUT has been calculated in \cite{Hisano:1996qq, Barbieri:1995tw}. A detectable signal is predicted in upcoming experiments, although the values considered for the sparticles masses are in tension with more recent LHC data \cite{WinoLHC}.

LFV can be sizeable in the context of GUT theories with right-handed sterile neutrinos, which has been studied in \cite{Barbieri:1995tw, Susyseesaw1,Susyseesaw2, Susyseesaw3, Casas:2001sr, Calibbi:2006nq, Calibbi:2012gr, Susyseesaw4, Susyseesaw5, Susyseesaw6}. In SO(10), right-handed neutrinos naturally appear in the $\textbf{16}$ spinor representation that an SM generation fills, and neutrino masses can be explained with a supersymmetric seesaw mechanism. Considering heavy right-handed neutrinos, the superpotential in the lepton sector reads
\begin{equation}
	W_{L}=[Y_e]_{ij}\bar{L}_i H_dE_j+[Y_\nu]_{ij} \bar{L}_i H_u N_j+\frac{1}{2}[M_R]_{ij} N_iN_j
\end{equation}  
where the notation for the SM superfields is self-explanatory, and $N$ is the superfield that contains sterile neutrinos. As in Equation~(\ref{eq:seesawformula}), for large Majorana masses $M_R$, the light neutrino mass matrix is
\begin{equation}
	m_\nu=-Y_\nu M^{-1}_R Y^T_\nu v^2\sin^2 \beta\qquad \text{where}\ \tan \beta=\frac{\expval{H_u}}{\expval{H_d}},\ v=174\ \text{G}e\text{V}
\end{equation}

The gravity-mediated soft breaking terms regarding sleptons are the following
\begin{equation}
	-\mathcal{L}_{ soft}=m^2_0(\tilde{\ell}^\dagger_{i}\tilde{\ell}_{i}+\tilde{\bar{e}}^\dagger_{i}\tilde{e}_{i})+m_0a_0([Y_e]_{ij}\tilde{\ell}^\dagger_{i}H_d\tilde{e}_{j}+[Y_\nu]_{ij}\tilde{\ell}^\dagger_{i}H_u\tilde{N}_{j}+\rm h.c)
\end{equation}
and the left-handed sleptons mass matrix is renormalized in the leading-log approximation as \cite{Susyseesaw1}
\begin{equation}
	[\Delta \tilde{m}^2_{L}]_{ij}= -\frac{1}{8\pi^2}[Y^\dagger_\nu Y_\nu]_{ij}m_0^2(3+\abs{a_0}^2)\log(\frac{M_{\rm Pl}}{M_{R}})
\end{equation}

The typical size of $l_i\to l_j \gamma$ branching fraction is \cite{Casas:2001sr}
\begin{equation}
    Br(l_i\to l_j\gamma)\sim \frac{\alpha^3_{em}}{G_F^2}\frac{\abs{\Delta \tilde{m}^2_{Lji}}^2}{m^8_{SUSY}}\tan^2 \beta\times Br(l_i\to l_j \bar{\nu}_j \nu_i)
\end{equation}
where $m_{SUSY}$ is the sparticles mass scale. In general, even knowing neutrino masses and mixing angles, the neutrino Yukawa $Y_\nu$ is not uniquely defined \cite{SachaIbarraGUTparameters}. In a basis where the Majorana masses $\hat{M}_R$ are diagonal, we can use the Casas-Ibarra parametrization \cite{Casas:2001sr} for $Y_\nu$
\begin{equation}
	Y_\nu\sim (U\sqrt{\hat{m}_\nu}R\sqrt{\hat{M}_R})/(v\sin\beta)^2
\end{equation}
where $U$ is PMNS and $R$ is an unknown orthogonal complex matrix. The matrix that controls the slepton 
 mixing is then
\begin{equation}
	Y^\dagger_\nu Y_\nu\sim \sqrt{\hat{M}_R} R^\dagger \hat{m}_\nu R \sqrt{\hat{M}_R}
\end{equation}
and depends on the unknown $R$ matrix. Assuming specific mass hierarchy and degenerate patterns for neutrinos, the free parameters in $R$ are reduced, and the predicted LFV signals are studied when the parameters are varied \cite{Masina:2005am}. In SO(10) GUT, the neutrino and up Yukawa are unified at the GUT scale, and different breaking scenarios can lead to lepton flavour change governed by CKM or PMNS mixing with the third generation \cite{Calibbi:2006nq, Calibbi:2012gr}. The PMNS angles are large and lead to an insufficiently suppressed $\mu\to e \gamma$ rate, larger than the current upper limit $Br(\mu\to e \gamma)<4.2\times 10^{-13}$ \cite{themegcollaboration2016search}. In the scenario where LFV amplitudes are proportional to CKM matrix elements, the rate is compatible with the future experimental upper bound $Br(\mu\to e \gamma)<6\times 10^{-14}$ \cite{meg_ii}. This model relates the branching ratios of $\tau\to l\gamma$ and $\mu \to e \gamma$ via the following relation
\begin{equation}
    Br(\tau\to \mu\gamma)\sim \frac{\abs{V_{33}V_{23}}^2}{\abs{V_{13}V_{23}}^2}Br(\mu\to e \gamma) \times 10^{-1}\lesssim 10^{-10}
\end{equation}
where $V$ is the CKM matrix, and we have substituted the upcoming MEG branching fraction bound. If a $\tau \to \mu\gamma$ signal is observed by BELLE II with $Br(\tau\to \mu\gamma)\sim 10^{-9}$ \cite{belle_ii}, it can disfavor the model. 
In the context of sleptons mixing and LFV, several simplified SUSY scenarios have been more recently studied, complemented with the bounds of the null results of the LHC \cite{Calibbi:2015kja}.

The soft-breaking sector is not the only possible source of LFV in supersymmetric SM. Gauge and SUSY invariance allow for the following terms in the superpotential:
\begin{equation}
    W_{RPV}=\frac{\lambda_{ijk}}{2} L_i L_j \bar{E}_k+\lambda'_{ijk}L_iQ_j\bar{D}_k+\lambda_{ijk}''\bar{U}_i\bar{D}_j\bar{D}_k+\mu_iL_i H_u \label{eq:RPviolating}
\end{equation}

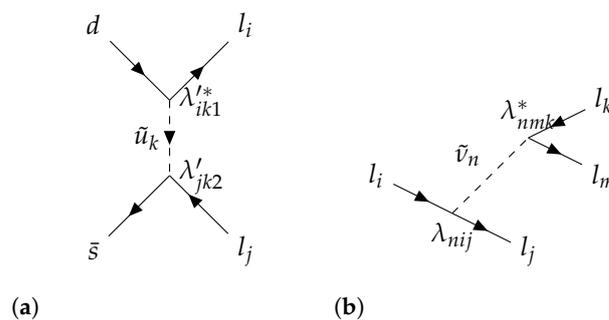
\begin{figure}[H]
\begin{subfigure}{.3\textwidth}
	\centering
\begin{tikzpicture}[scale=1]
	\begin{feynman}[small]
		\vertex (particle1) at (-1,2) {\(d\)};
		\vertex (particle2) at (1,2) {\(l_i\)};
		\vertex (particle3) at (-1, -1) {\(\bar{s}\)};
		\vertex (particle4) at (1, -1) {\(l_j\)};
		\vertex (vertex12) at (0,1) [label=0:\(\lambda'^{*}_{ik1} \)];
		\vertex (vertex34) at (0,0) [label=0:\(\lambda'_{jk2} \)];
		\diagram* [inline=(a.base)]{
			(vertex12) -- [charged scalar, edge label'=\(\tilde{u}_k\)] (vertex34),
			(particle1) -- [fermion] (vertex12) -- [fermion] (particle2),
            (particle3) -- [anti fermion] (vertex34) -- [anti fermion] (particle4),
		};
	\end{feynman}
\end{tikzpicture}
\caption{}\label{subfig:RPVMeson}
\end{subfigure}
\begin{subfigure}{.3\textwidth}
	\centering
\begin{tikzpicture}[scale=1]
	\begin{feynman}[small]
		\vertex (particle1) at (-1,0.5) {\(l_i\)};
		\vertex (particle2) at (1,-0.5) {\(l_j\)};
		\vertex (particle3) at (2, 1.5) {\(l_k\)};
		\vertex (particle4) at (2, 0.5) {\(l_m\)};
		\vertex (vertex12) at (0,0) [label=-90:\(\lambda_{nij} \)];
		\vertex (vertex34) at (1,1) [label=90:\(\lambda^*_{nmk} \)];
		\diagram* [inline=(a.base)]{
			(vertex12) -- [scalar, edge label=\(\tilde{\nu}_n\)] (vertex34),
			(particle1) -- [fermion] (vertex12) -- [fermion] (particle2),
            (particle3) -- [fermion] (vertex34) -- [fermion] (particle4),
		};
	\end{feynman}
\end{tikzpicture}
\caption{}\label{subfig:RPV3l}
\end{subfigure}
\caption{Examples of LFV tree-level diagrams in the supersymmetric SM with R-parity violation. ({\bf a}) Diagrams contributing to the LFV meson decay $K^0\to l_i l_j$.
({\bf b}) Diagrams contributing to the $l_i\to l_j l_k l_m$ rate. }\label{fig:RPVLFV}
\end{figure}

The $\lambda''$ term is baryon number violating and can lead to prompt proton decay. To avoid this disastrous outcome, a discrete symmetry known as $R-$parity is often assumed. The $R-$parity of a particle is defined as $(-1)^R\equiv(-1)^{3(B-L)+2S}$, where $B,L$ are the baryon and lepton number, while $S$ is the particle spin. It follows that any SM particle is $RP-$even and the superpartners are $RP-$odd. $RP-$invariance automatically forbids all terms in the superpotential of Equation~(\ref{eq:RPviolating}), but other discrete symmetries such as baryon parity \cite{IBANEZ19923} can allow for lepton flavor violation while conserving baryon number. The first two terms in the superpotential leads to the Lagrangian terms 
 \cite{Rpreview}
\begin{align}
    \mathcal{L}_{RPV}=\lambda_{ijk}(\overline{\nu^c}_{Li} e_{Lj} \tilde{e}_{Rk}^\dagger+ \bar{e}_{Rk} \nu_{Li}\tilde{e}_{Lj}+\bar{e}_{Rk}e_{Lj}\tilde{\nu}_{iL})\nonumber \\
    +\lambda'_{ijk}(V_{jm}\bar{d}_{Rk}d_{Lm}\tilde{\nu}_{iL}+V_{jm}\bar{d}_{Rk}\nu_{Li}\tilde{d}_{mL}+V_{jm}\overline{\nu^c}_{Li}d_{Lm}\tilde{d}^\dagger_{Rk}+\nonumber \\-\bar{d}_{Rk}u_{Lj}\tilde{e}_{Li}-\bar{d}_{Rk}e_{Li}\tilde{u}_{Lj}-\overline{e^c}_{Li}u_{Lj}\tilde{d}^\dagger_{Rk})+ \rm{h.c}
\end{align}
that can allow for several LFV processes already at tree level. In Figure \ref{fig:RPVLFV}a, we show a diagram for the LFV $K^0$ decay $K^0_L\to\mu e$, whose branching fraction is constrained by the current upper limit $Br(K^0_{L}\to \mu e)<4.7\times 10^{-12}$ \cite{bnl_k0l}. Assuming only one non-zero pair of R-parity violating coupling $\lambda'^{*}_{ik1}\lambda'_{jk2}$, the bound implies (adapted from \cite{RPVLFVMeson})
\begin{equation}
    \abs{\lambda'^{*}_{1k1}\lambda'_{2k2}}\times \left(\frac{100\ \text{G}e\text{V}}{m_{\tilde{u}_k}}\right)^2< 1.3 \times 10^{-7}\qquad \rightarrow\qquad \abs{\lambda'^{*}_{1k1}\lambda'_{2k2}}\lesssim 10^{-4} 
\end{equation}
\textls[-25]{where we have assumed $m_{\tilde{u}_k}\sim \rm{few}\times\ \text{T}e\text{V}$. Similarly, $Br(\mu\to 3e)<10^{-12}\to 10^{-16}$~\mbox{\cite{mu3e, mu3e_2}}} can set the following constraint on the coupling products $\abs{\lambda_{n21}\lambda^*_{n11}}$ ($\lambda$ is anti-symmetric in the first two indices and $n\neq 1$) from the diagrams of Figure \ref{fig:RPVLFV}b:
\begin{equation}
    \abs{\lambda_{n21}\lambda^*_{n11}}\times \left(\frac{100\ \text{Ge}\text{V}}{m_{\tilde{\nu}_n}}\right)^2<6.6\times 10^{-7}\ (6.6\times 10^{-9})
\end{equation}
$\lambda \lambda'$ diagram give tree-level contributions to $\mu\to e$ conversion, and at one loop $l_i\to l_j \gamma$
is sensitive to $\lambda\lambda, \lambda'\lambda'$ couplings. For a more complete discussion on LFV in R-parity violating theories, we refer the reader to \cite{RPVLFV1, RPVLFV2, RPVLFV3Baryons, RPVLFV4mutoeconversion, RPV5, RPV6}.
\vspace{-9pt}

\subsection{Effective Field Theory for Charged Lepton Flavor Violation}
Under the assumption that the new physics responsible for CLFV is heavy, i.e., $\Lambda>$ T$e$V, Effective Field Theories (EFT) are powerful model-independent descriptive frameworks. In EFT, the UV physics is hidden in the Wilson Coefficients (WC) of non-renormalizable interactions among light degrees of freedom that are suppressed by inverse power of the new physics scale
\begin{equation}
    \mathcal{L}_{\rm EFT}=\mathcal{L}_{d=4}+\sum_{I,d>4} \frac{C_d^I\mathcal{O}^I_d}{\Lambda^{d-4}}
\end{equation}

For a general introduction on EFT, see \cite{GeorgiEFT,EFTLesHouches}. In a $"$top-down$"$ perspective, the heavy fields of a particular model can be integrated out, and a tower of higher dimensional operators is generated, whose coefficients are expressed in terms of the couplings and parameters of the model. In this approach, the EFT plays the role of a helpful calculational tool, since observables are more easily computed with contact interactions constructed out of light fields rather than in the full theory. In a ``bottom-up$"$ analysis, rates are calculated in the most general EFT that is relevant for the processes considered, and bounds or observations are translated to the operator coefficients, which allow to probe general heavy physics model. Bottom-up calculations should include every contribution to which observables can be sensitive.

\textls[-15]{At fixed order in the EFT expansion in inverse powers of the heavy scale, EFT renormalization proceeds as in any field theory, and the couplings run with the energy scale. Observables are computed in terms of WC at the experimental energy scale that, for the many LFV channels, lies below the electroweak scale. 
The EFT permits the separation of energy scales and is most helpful when appropriate for the process considered: it should include the dynamical degrees of freedom and respect the symmetries that concern the probed energy scale. Below the electroweak scale, the top quark, the Higgs boson and the electroweak $W$, $Z$ boson can be integrated out and the EFT features contact interactions among lighter SM particles, respecting QED and QCD gauge symmetries. We reproduce in Table \ref{tab:EFTbasis} the operator basis of \cite{SachaCC} relevant for $\mu\to e$ LFV processes that are otherwise flavour diagonal. }
\begin{table}[H]
\caption{Low-energy QCD$\otimes$QED invariant EFT for $\mu\to e$, $\mu\to 3e$ and $\mu \to e$ conversion in nuclei, in the notation of 
 \cite{SachaCC}. The experiments can be sensitive to three, four point functions that correspond to operators of dimension from five to eight. $X_{\alpha\beta}=F_{\alpha\beta},G_{\alpha\beta}$ are the field tensors of photon and gluons, respectively. The chiral projector $P_{Y,Z}$ can be $Y,Z\in \{L,R\}$, $\bar{L}=R,\bar{R}=L$, while $q\in \{u,d,s,b,c\}$ and $l\in \{e,\mu,\tau\}$.}
  \begin{tabularx}{\textwidth}{m{5cm}<{\centering}m{7cm}<{\centering}}
\toprule
   \multicolumn{2}{c}{$2l$ operators}\\
       \midrule
\addlinespace[1ex]
$\mathcal{O}_{D,Y}$ & $\frac{m_\mu}{\Lambda^2}(\bar{e} \sigma^{\alpha\beta}P_Y\mu) F_{\alpha\beta}$\\
$\mathcal{O}_{XX,Y}$ & $\frac{1}{\Lambda^3}(\bar{e} P_Y \mu) X_{\alpha\beta}X^{\alpha\beta}$\\
$\mathcal{O}_{X\tilde{X},Y}$ & $\frac{1}{\Lambda^3}(\bar{e} P_Y \mu) X_{\alpha\beta}\tilde{X}^{\alpha\beta}$\\
$\mathcal{O}_{XXV,Y}$ & $\frac{1}{\Lambda^4}(\bar{e} \gamma_\sigma P_Y \mu) X_{\alpha\beta}\partial_\beta X^{\alpha\sigma}$\\
$\mathcal{O}_{X\tilde{X}V,Y}$ & $\frac{1}{\Lambda^4}(\bar{e} \gamma_\sigma P_Y \mu) X_{\alpha\beta}\partial_\beta \tilde{X}^{\alpha\sigma}$\\
\midrule
   \multicolumn{2}{c}{$2l2q$ operators}\\
       \midrule
$\mathcal{O}^{qq}_{V,YZ}$ & $\frac{1}{\Lambda^2}(\bar{e} \gamma^\alpha P_Y\mu)(\bar{q} \gamma_\alpha P_Z q)$\\
$\mathcal{O}^{qq}_{S,YZ}$ & $\frac{1}{\Lambda^2}(\bar{e} P_Y\mu)(\bar{q} P_Z q)$\\
$\mathcal{O}^{qq}_{T,YY}$ & $\frac{1}{\Lambda^2}(\bar{e} \sigma^{\alpha\beta}P_Y\mu)(\bar{q} \sigma_{\alpha\beta}P_Y q)$\\
\midrule
 \multicolumn{2}{c}{$4l$ operators}\\
       \midrule
$\mathcal{O}^{ll}_{V,YZ}$ & $\frac{1}{\Lambda^2}(\bar{e} \gamma^\alpha P_Y\mu)(\bar{l} \gamma_\alpha P_Z l)$\\
$\mathcal{O}^{ll}_{S,YY}$ & $\frac{1}{\Lambda^2}(\bar{e} P_Y\mu)(\bar{l}  P_Y l)$\\
$\mathcal{O}^{\tau\tau}_{S,Y\bar{Y}}$ & $\frac{1}{\Lambda^2}(\bar{e} P_Y\mu)(\bar{\tau}  P_{\bar{Y}} \tau)$\\
 $\mathcal{O}^{\tau\tau}_{T,YY}$ & $\frac{1}{\Lambda^2}(\bar{e} \sigma^{\alpha\beta}P_Y\mu)(\bar{\tau} \sigma_{\alpha\beta} P_Y \tau)$\\
\bottomrule
\end{tabularx}
\label{tab:EFTbasis}
\end{table}
\vspace{-6pt}
Normalizing the operators of Table \ref{tab:EFTbasis} at $\Lambda=v$, where $v^2=1/(2\sqrt{2}G_F)$, the branching fraction of $\mu\to e\gamma $ in the EFT is \cite{Kuno:1999jp}
\begin{equation}
    Br(\mu\to e \gamma)=384\pi^2(\abs{C_{D,L}}^2+\abs{C_{D,R}}^2)<4.2\times 10^{-13}\to \abs{C_{D,Y}}<1.05\times 10^{-8} \label{eq:mutoegammaEFT}
\end{equation}
where the dipole operator coefficient $C_{D,Y}(M)$ are at $M=m_\mu$. If we assume that the dipole coefficient is order one if normalized at the new physics scale, Equation~(\ref{eq:mutoegammaEFT}) implies $\Lambda_{\rm NP}\sim 1.6\times 10^4 $ TeV. The bound can be satisfied with lighter UV physics if the dipole coefficient is loop and/or coupling suppressed.

\textls[-15]{To use the bound on EFT operator coefficients to constrain generic BSM heavy physics, we should determine the upper limit on coefficients at the new physics scale, where the heavy degrees of freedom are integrated out. This is done by solving the renormalization group equations of the Wilson coefficients, which requires dressing the operator basis with QED and QCD loops. QED amounts to a few percent effect in the renormalization of operator coefficients, but it still plays an important role because it can mix operators. QCD running does not mix operators, but the rescaling of quark scalar and tensors coefficients is numerically relevant (few$\times 10\%$). Operator mixing allows probing an operator coefficient which is difficult to detect via its mixing to a tightly constrained one. For instance, the tensor operator $\mathcal{O}^{\tau\tau}_{T,Y}=(\bar{e} \sigma^{\alpha\beta}P_Y\mu)(\bar{\tau} \sigma_{\alpha\beta} P_Y \tau)$ mixes into the dipole by closing the tau legs in a loop and attaching a photon. The contribution to the dipole coefficient is one-loop suppressed (and log enhanced), but, to close the loop, a chirality flip is necessary, and a $\tau$ mass insertion enhances the mixing by $m_\tau/m_\mu$. Complemented with a large anomalous dimension, the mixing is $\sim \mathcal{O}(1)$. The (sensitivity) bound that $\mu\to e \gamma$ sets on the tensor coefficient at $m_W$ is then \cite{SachaCC}}
\vspace{-6pt}
\begin{equation}
    C^{\tau\tau}_{T,Y}(m_W)\lesssim 1.07\times 10^{-8} \label{eq:boundtensormutoegamma}
\end{equation}

We should stress that this is not an exclusion bound, but rather an experimental sensitivity. In coefficient space, $\mu\to e \gamma$ constrains one direction that corresponds to the dipole coefficient at the experimental scale $m_\mu$. The RGEs can tell us how this direction rotates in the coefficient space at higher energies, but the experiment still imposes a bound in one direction only. In other words, the bound will apply to a single combination of operator coefficients at the high scale; namely, solving the RGEs up to $m_W$, 
\vspace{-6pt}
\begin{equation}
    \abs{C_{D,Y}}(m_\mu)=\abs{0.938 C_{D,Y}(m_W)+0.981 C^{\tau\tau}_{T,Y}(m_W)+\dots}<1.05\times 10^{-8}. \label{eq:runneddipole}
\end{equation}

The upper limit in Equation~(\ref{eq:boundtensormutoegamma}) corresponds to the case where only $C^{\tau\tau}_{T,Y}(m_W)$ is non-zero and is commonly known as one-operator-at-a-time sensitivity. A sensitivity corresponds to the smallest absolute value which is experimentally detectable, but larger values are possible if cancellations with other contributions occur. Indeed, we can see that if the dipole and the tensor are of similar size and opposite sign, an accidental cancellation can occur in Equation~(\ref{eq:runneddipole}). This is an example of a flat direction in coefficient space. Flat directions are a general feature of bottom-up EFT analyses of LFV, because the operator basis contains more operators than observables, and the few operators constrained by experiments mix with the rest in the RGEs. Nonetheless, identifying operator coefficients to which observables are most sensitive is a useful guide for model building.  The sensitivities of $\mu\to 3e, \mu\to e\gamma$ and $\mu N\to e N$ to Wilson coefficients at $m_W$ in the low energy EFT has been extensively studied~\cite{SachaCC, SachaEFT1, SachaEFT2, SachaEFT3, SachaEFT4}. Spin-dependent $\mu\to e$ conversion in nuclei~\mbox{\cite{SachaEFT2, SachaEFT3, Hoferichter:2022mna, SachaDiffTargets}}, although less constraining than the spin-independent searches, allows us to probe different combinations of coefficients and, thus, reduce the number of flat directions. 

Leptonic and semi-leptonic rare meson decays such as $K^0_L\to \mu e$, $K^+\to \pi^+ \mu e$ are systematically studied in the EFT by adding to the operator basis quark flavor-changing operators, and the sensitivities to Wilson coefficient can be similarly determined \cite{SachaMesonDecay}.  Concerning $\tau\leftrightarrow l$ processes, EFT analysis can be found in \cite{SachaCC, Husek:2020fru}. 

\textls[-15]{Above the electroweak scale, the appropriate EFT is the Standard Model Effective Field Theory (SMEFT), which contains all SM particles, and the operators are $SU(3)\otimes SU(2)\otimes U(1)$ invariant. The contact interactions in the low energy EFT for LFV are matched onto SMEFT operators of dimensions between six and eight \cite{Dimension6SMEFT, Dimension7SMEFT, Dimension8SMEFT1, Dimension8SMEFT2}. Running the experimental bound from $m_W$ to a higher scale requires solving the RGEs of the SMEFT operators~\mbox{\cite{RGEsdimsix1, RGEsdimsix2, RGEsdimsix3}}, identifying the contributions to which LFV observables are sensitive. Establishing all the relevant contributions in SMEFT is not trivial, and a complete bottom-up analysis of LFV  requires some challenging calculations that are currently missing in the literature \cite{Ardu:2021koz}.}

\section{Experimental Review}\label{sec:exp}
Searches for evidence of CLFV signals span a broad range of experimental techniques thanks to the large variety of processes one could be looking for, such as rare muon and tau decays (\meg, \mueee, \taumug, \taueg, \taulll), rare mesons and bosons decays, and direct conversions of a lepton in a nuclear field (\muconv, \muconvp). Table~\ref{tab:clfv_limits} summarizes the current best limits on the various channels.

\begin{table}[H]
  \centering
  \small
  \caption{
    \label{tab:clfv_limits}
    Current experimental upper limits on the branching ratios of CLFV processes for muons, taus, mesons ($\pi$, $J/\psi$, $B$, $K$) and bosons ($Z^0$, $h$).
  }
\begin{tabularx}{\textwidth}{m{3.5cm}<{\centering}m{2cm}<{\centering}m{5cm}<{\centering}m{2cm}<{\centering}}
 
    \toprule
    \textbf{Process} & \textbf{Experiment} & \textbf{Limit} & \textbf{C.L.}    \\
    \midrule
    \meg          & MEG         & $4.2\times 10^{-13}$~\cite{themegcollaboration2016search} & 90\%\\
    \midrule
    \mueeep       & SINDRUM     & $1.0\times 10^{-12}$~\cite{BELLGARDT19881} & 90\%\\ 
    \midrule
    \muconv       & SINDRUM-II  &  $6.1(7.1)\times 10^{-13}$ Ti (Au)~\cite{DOHMEN1993631,SINDRUM_II}& 90\%\\
    \midrule
    \muconvp      & SINDRUM-II  &  $5.7\times 10^{-13}$~\cite{1998334}& 90\%\\
    \midrule
    
    \taueg        & BaBar  &  $3.3\times 10^{-8}$~\cite{babar_taulg}& 90\%\\
    \midrule
    \taumug       & BaBar  &  $4.4\times 10^{-8}$~\cite{babar_taulg}& 90\%\\
    \midrule
    $\tau\to eee$       & Belle  &  $2.7\times 10^{-8}$~\cite{belle_taulll}& 90\%\\
    \midrule
    $\tau\to \mu\mu\mu$ & Belle  &  $2.1\times 10^{-8}$~\cite{belle_taulll}& 90\%\\
    \midrule
    $\tau\to \mu e e$   & Belle  &  $1.8\times 10^{-8}$~\cite{belle_taulll}& 90\%\\
    \midrule
    $\tau\to e\mu\mu$   & Belle  &  $2.7\times 10^{-8}$~\cite{belle_taulll}& 90\%\\
    \midrule
    $\tau\to \pi^0 e$   & Belle  &  $8.0\times 10^{-8}$~\cite{belle_pil}& 90\%\\
    \midrule
    $\tau\to \pi^0 \mu$ & BaBar  &  $1.1\times 10^{-7}$~\cite{babar_pil}& 90\%\\
    \midrule
    $\tau\to \eta e$   & Belle  &  $9.2\times 10^{-8}$~\cite{belle_pil}& 90\%\\
    \midrule
    $\tau\to \eta \mu$ & Belle  &  $6.5\times 10^{-8}$~\cite{belle_pil}& 90\%\\
    \midrule
    $\tau\to \rho^0 e$  & Belle  &  $1.8\times 10^{-8}$~\cite{belle_rhol}& 90\%\\
    \midrule
    $\tau\to \rho^0 \mu$& Belle  &  $1.2\times 10^{-8}$~\cite{belle_rhol}& 90\%\\
    \midrule
    
    $\pi^0\to \mu e$                & KTeV  &  $3.6\times 10^{-10}$~\cite{ktev_limits}& 90\%\\
    \midrule
    $K^0_L\to \pi^0 \mu^+ e^-$      & kTeV  &  $7.6\times 10^{-11}$~\cite{ktev_limits}& 90\%\\
    \midrule
    $K^0_L\to \mu e$          & BNL E871  &  $4.7\times 10^{-12}$~\cite{bnl_k0l}& 90\%\\
    \midrule
    $K^+\to \pi^+ \mu^+ e^-$  & BNL E865  &  $1.3\times 10^{-11}$~\cite{bnl_kpimue}& 90\%\\
    \midrule
    $J/\psi \to \mu e$       & BESIII  &  $1.5\times 10^{-7}$~\cite{besiii_jpsiemu}& 90\%\\
    \midrule
    $J/\psi \to \tau e$      & BESIII  &  $7.5\times 10^{-8}$~\cite{besiii_jpsietau}& 90\%\\
    \midrule
    $J/\psi \to \tau \mu$    & BESII  &  $2.6\times 10^{-6}$~\cite{besii_jpsitaumu}& 90\%\\
    \midrule
    $B^0 \to \mu e$       & LHCb  &  $2.8\times 10^{-9}$~\cite{lhcb_b0mue}& 95\%\\
    \midrule
    $B^0 \to \tau e$      & BaBar  &  $2.8\times 10^{-5}$~\cite{babar_b0taul}& 90\%\\
    \midrule
    $B^0 \to \tau \mu$    &  LHCb &  $1.4\times 10^{-5}$~\cite{lhcb_b0taumu}& 95\%\\
    \midrule        
    $B \to K \mu e$       & BaBar  &  $3.8\times 10^{-8}$~\cite{babar_kll}& 90\%\\
    \midrule
    $B \to K^* \mu e$     &  BaBar &  $5.1\times 10^{-7}$~\cite{babar_kll}& 90\%\\
    \midrule
    $B^+ \to K^+\tau e$   & BaBar  &  $4.8\times 10^{-5}$~\cite{babar_ktaul}& 90\%\\
    \midrule
    $B^+ \to K^+\tau \mu$ & BaBar  &  $3.0\times 10^{-5}$~\cite{babar_ktaul}& 90\%\\
    \midrule        
    $B^0_s \to \mu e$     & LHCb  &  $1.1\times 10^{-8}$~\cite{lhcb_b0mue}& 90\%\\
    \midrule        
    $B^0_s \to \tau\mu$   & LHCb  &  $4.2\times 10^{-5}$~\cite{lhcb_b0taumu}& 95\%\\
    \midrule
    $Z^0 \to \mu e$       & ATLAS  &  $7.5\times 10^{-7}$~\cite{atlas_zmue}& 95\%\\
    \midrule
    $Z^0 \to \tau e$      & OPAL  &  $9.8\times 10^{-6}$~\cite{opal_ztaue}& 95\%\\
    \midrule
    $Z^0 \to \tau \mu$    & DELPHI  &  $1.2\times 10^{-5}$~\cite{delphi_ztaumu}& 95\%\\
    \midrule        
    $h \to \mu e$         & ATLAS  &  $6.1\times 10^{-5}$~\cite{ATLAS:2019xlq}& 95\%\\
    \midrule
    $h \to \tau e$        & CMS  &  $2.2\times 10^{-3}$~\cite{cms_htaul}& 95\%\\
    \midrule
    $h \to \tau \mu$      & CMS  &  $1.5\times 10^{-3}$~\cite{cms_htaul}& 95\%\\
    \bottomrule
  \end{tabularx}
\end{table}

The CLFV searches based on muons have been performed with dedicated experiments, usually highly tuned for a specific channel, which took advantage of the facilities capable of delivering a high intensity muon beam (see next section). For all the other cases (tau, mesons and bosons), with the only exception made for the Kaons, it is not possible to deliver a dedicated beam; thus, general-purpose detector systems have been used. 

As discussed in the previous sections, the most stringent constraints 
 on various BSM models are set by the direct searches of CLFV decays of muons and taus decays. 
In the following sections, we describe the most recent and the coming experimental efforts for these two categories: (i) searches using muons, (ii) searches using taus. For each search, a discussion of the peculiarities of the signal topology and of the various experimental challenges is provided. 

\subsection{CLFV Searches Using Muons}\label{sec:muon_sector}
In the history of CLFV experiments, muons have been, so far, the most popular. Historically, the first experiment looking for CLFV using muons was performed by Hinks and Pontecorvo using atmospheric muons~\cite{PhysRev.73.257}. Since then, the advancements in the muon beam production/acceleration technology at different facilities (PSI, TRIUMPH, LANL, etc.) made available high-intensity muon beams at the level of $10^{8}(10^{7}) \mu^+(\mu^-)/s$~\cite{Signorelli,10.1007/3-540-30924-1_129}, enabling the possibility to search for rare CLFV processes. 
Facilities under construction at Fermilab (USA) and J-PARC (Japan)~\cite{Bernstein}, or planned, like the High Intensity Muon Beam project at the PSI (Switzerland)~\cite{aiba2021science}, have been designed to provide muon beams with an intensity of about 10$^{10}$ \textmu/s. This planned intensity corresponds to 2--3 orders of magnitude improvement with respect to the current state-of-the-art technology. 
The J-PARC and Fermilab muon experiments will use a novel method for creating the muon beam. A prototype muon beamline, the Muon Science Innovative Channel (MuSIC), was set up at the Research Center for Nuclear Physics (Osaka, Japan) to prove the conceptual idea. The production of an intense muon beam relies on the efficient capture of pions (from proton-target interactions), which subsequently decay to muons, using a novel superconducting solenoid magnet system~\cite{music}.

The current best limits on the muon-CLFV processes come from experiments that performed dedicated searches for the following processes: \meg, \mueee, \muconv and \muconvp. Table~\ref{tab:clfv_limits} summarizes these results. One thing to notice is that all the searches, except the \muconv and \muconvp, were performed using $\mu^+$ rather than $\mu^-$. This choice is motivated by several advantages: (i) $\mu^+$ cannot get captured in nuclei, while $\mu^-$ can undergo nuclear capture events, which produce protons, neutrons, photons and, thus, increase the activity in the detector deteriorating its performance, (ii) the muon beam is obtained from charged pions decay, which are produced in proton-target interactions where $\pi^+$ production is larger; thus, the resulting $\mu+$ beam intensity is higher.

The following sections offer a more detailed description of each of these experimental~searches.

\subsubsection{\meg}
In the \meg decay, the final state consists of a back-to-back positron and photon with an energy of 52.8 MeV.
The background sources for this search can be factorized into two main categories: (i) an intrinsic physics background from the Radiative Muon Decay (RMD) process $\mu^+\to e^+ \gamma \nu_e \bar{\nu}_\mu$, where the neutrinos carry off small momenta, and (ii) an “accidental” background where a positron from the Michel decay $\mu^+\to e^+\nu_e \bar{\nu}_\mu$, together with a photon from an RMD event or an electron-positron annihilation in flight or an $e-N$ nucleus scattering, recreate the topology of the \meg decay.
While signal and RMD rates are proportional to the muon stopping rate $R_\mu$, the accidental background rate is proportional to $R_\mu^2$ because both particles come from the beam; the accidental background is, therefore, the dominant enemy of this search. Thus, a continuous muon beam is better suited than a pulsed beam to avoid stripping particles in short bunches, and $R_\mu$ must be carefully chosen to optimize the sensitivity. 
The \meg searches from the last decades confirmed that the accidental background is dominant, while the intrinsic background accounts for about 10\% of the total background budget.
%

Two different strategies have been applied for designing the experimental apparatus for the \meg search: (i) a tracking-only system equipped with a converter to convert the photon in an $e^+e^-$ pair, or (ii) a tracker combined with a calorimeter for the photon detection. The tracking-only solution has a much better resolution but a cost of a loss in acceptance because converting the photon requires material that spoils the resolution (due to energy loss and multiple scattering) but too little limits the size of the data sample. 

\begin{figure}[H]
     \includegraphics[width =0.45\textwidth]{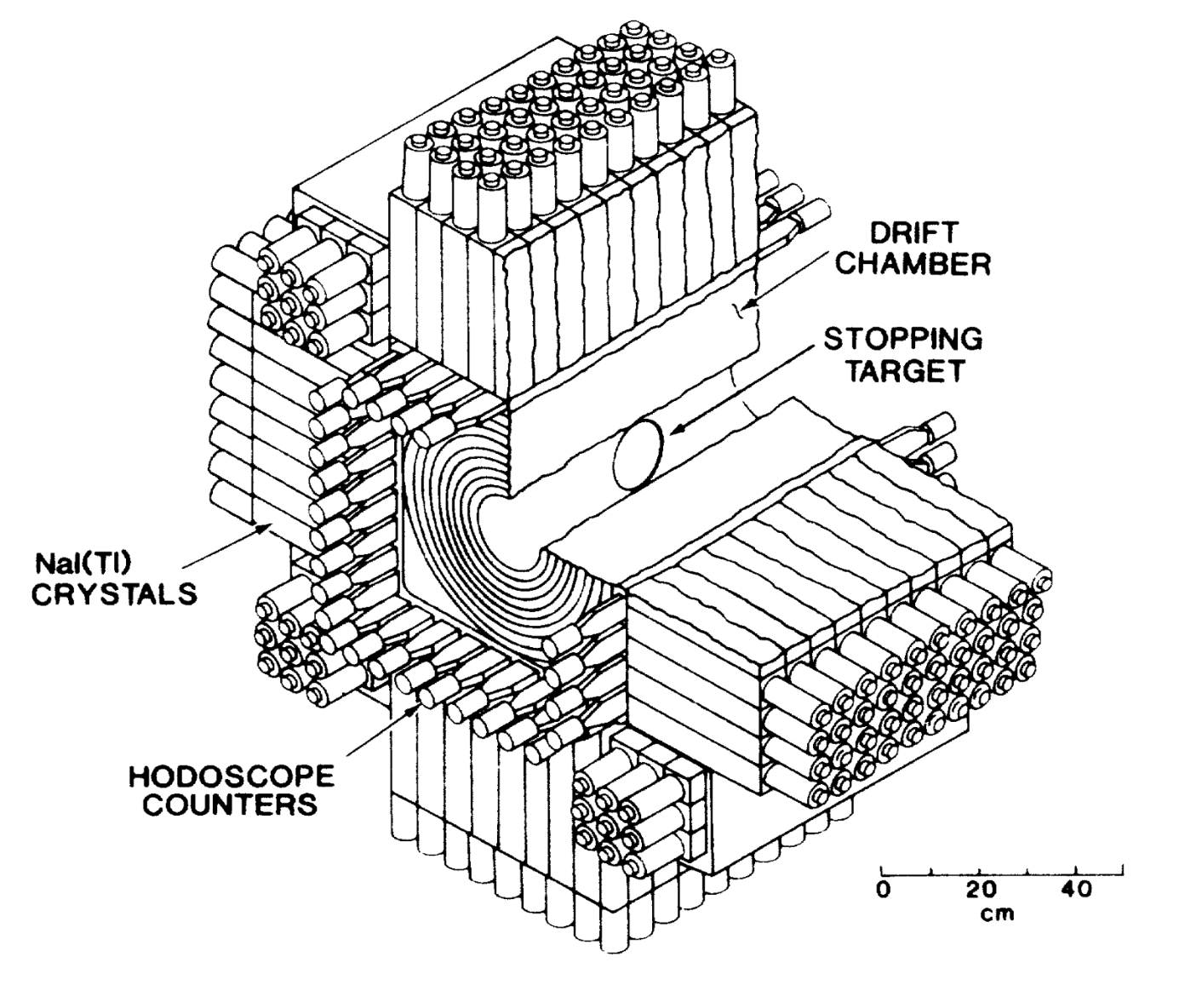}
     \caption{Schematic of the Crystal Box experiment (Figure from~\cite{crystalbox}).}\label{fig:CRYSTALBOX}
\end{figure}

One of the first experiments to adopt the calorimetric solution for the photon detection was the Crystal Box experiment at  Los Alamos Meson Physics Facility (LAMPF)~\cite{crystalbox}. The experiment, shown in Figure~\ref{fig:CRYSTALBOX}, used a surface muon beam at LAMPF with an average intensity of 300 kHz.  The detector consists of a cylindrical drift chamber surrounded by 396 NaI(Tl) crystals. A layer of scintillation counters in front of the crystals provided a timing measurement for the electrons and a veto for photons. The energy resolution for electrons and photons was $\sim$6\% (FWHM). The position resolution of the drift chamber was 350 \textmu m, while the time resolution was $\sim$400 ps for the scintillators and $\sim1$ ns for the crystals. 
A total of $3\times10^{12}$ muon were stopped in a thin polystyrene stopping target. A~maximum likelihood analysis established a 90\% C.L. upper limit of $4.9 \times 10^{-11}$~\cite{crystalbox}.

\begin{figure}[H]
     \includegraphics[width =0.45\textwidth]{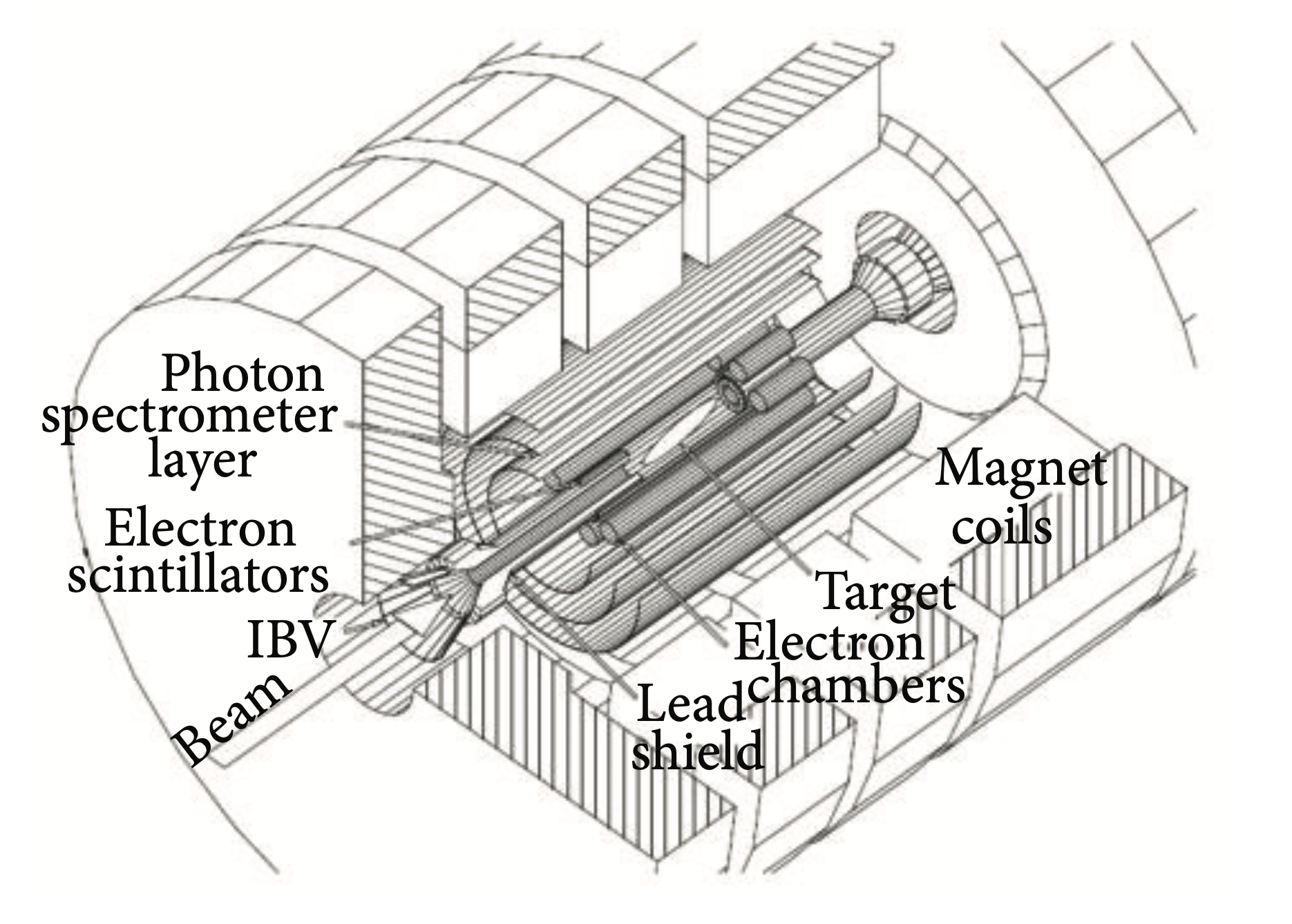}
     \caption{Schematic of the MEGA experiment (Figure from
~\cite{MEGA}).}\label{fig:MEGA}
\end{figure}

The next-generation experiment, MEGA~\cite{MEGA}, was also performed at LAMPF. The MEGA experimental apparatus, shown in Figure~\ref{fig:MEGA}, used a surface muon beam at the stopped muon channel at LAMPF that was stopped in a 76 \textmu m Mylar foil centered in the 1.5 T magnetic field of a superconducting solenoid. 
The MEGA detector consisted of a magnetic spectrometer for the positron and three spectrometers for the photon, therefore sacrificing the signal acceptance and efficiency for a better resolution and background rejection. In total, $1.2\times10^{14}$ muons were stopped during the life of the experiment, and the overall efficiency for the observation of the \meg event was $\sim$3.9 $\times$ 10$^{-3}$. The small efficiency was due to the photon conversion probability ($\sim$2.5\%) and to the reduced capability of reconstructing the  positron tracks in the solenoidal field compared to the design value. For these reasons, the final sensitivity reached by the MEGA experiment, $1.2 \times 10^{-11}$ @ 90\% C.L.~\cite{MEGA}, was $\sim$35 times worse than the design value, proving how challenging it is to deliver progress in this type of search.

\begin{figure}[H]
     \includegraphics[width =0.75\textwidth]{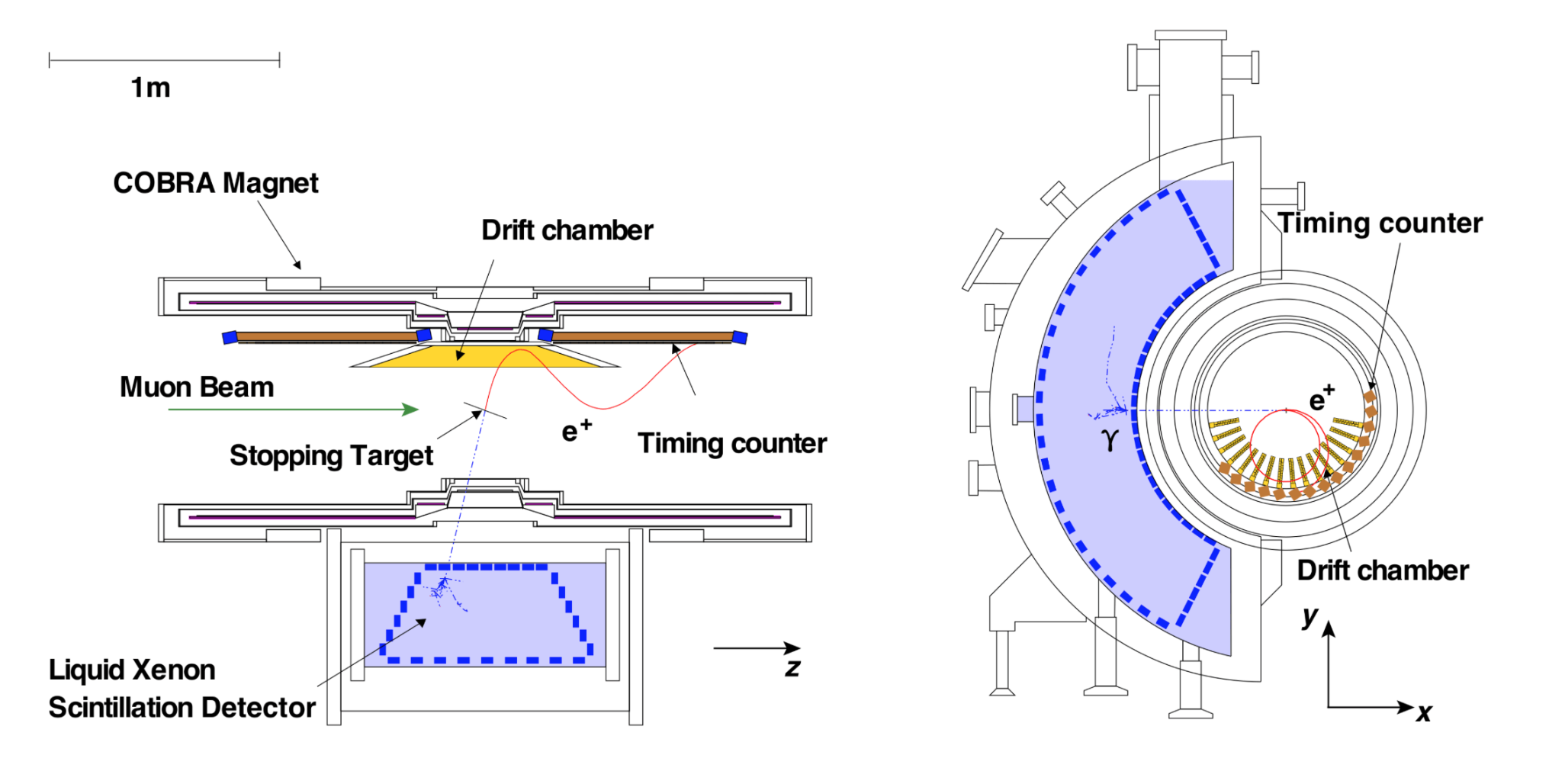}
     \caption{Schematic of the MEG experiment (Figure from
~\cite{themegcollaboration2016search}).}\label{fig:MEG}
\end{figure}

The current best limit for the \meg branching ratio, $4.2\times 10^{-13}$ @ 90\% C.L., comes from the MEG experiment~\cite{themegcollaboration2016search}. The detector system, shown in Figure~\ref{fig:MEG}, covers  $\sim$10\% of the solid angle and surrounds a 205 \textmu m-thick polyethylene muon stopping target. The apparatus consists of a positron spectrometer and a liquid-xenon (LXe) calorimeter. 

MEG opted for no converter for the photon detection, the opposite of MEGA. This choice avoids the pileup problem in the pattern recognition that limited MEGA but, at the same time, limits the geometrical acceptance. Table~\ref{tab:meg_perf} summarizes the detector performance measured during the MEG operation~\cite{meg_ii}.
A key feature of MEG is the magnetic field design. MEG adopted a graded solenoidal field, set at $\sim$1.1 T near the center of the apparatus, that sweeps out the positrons emitted at $\sim$90 $\deg$ and provides a constant bending radius for the signal positron essentially independent of the angle of emission. This feature helps in achieving a uniform and efficient signal track reconstruction. 
Another technological breakthrough from the MEG experiment is the development of the liquid Xe (LXe) calorimeter. The MEG LXe calorimeter is the first application of a large volume of LXe for particle detection and, so far, it proved to have the best performance for the electromagnetic calorimetry detection in the energy range below 100 MeV~\cite{meg_lxe}.

\vspace{-12pt} 


\vspace{-9pt}

\begin{table}[H]
  \small
  \caption{
    \label{tab:meg_perf}
    Summary of detector performance for the MEG and MEG-II experiments~\cite{meg_ii}. $\sigma_X$ indicates the resolution of the observable $X$,  $\epsilon_X$ the detection efficiency for the  particle $X$. For the case of the photon energy resolution $\sigma_{E_\gamma}$, the two values refer to the shallow ($<$2 cm)/deep ($>$2 cm) events. $\sigma_{t_{e^+\gamma}}$ is the time  resolution on the $e^+-\gamma$ time residual. The reported values for the MEG-II case refer to the updated results from the engineering runs reported in
~\cite{meg_ii}.
  }
\begin{tabularx}{\textwidth}{m{2cm}<{\centering}m{1.5cm}<{\centering}m{2cm}<{\centering}m{1.5cm}<{\centering}m{1.5cm}<{\centering}m{1.5cm}<{\centering}m{1.5cm}<{\centering}m{1.5cm}<{\centering}}
    \toprule
                  & \boldmath$\sigma_{p_e^+}$ &  \boldmath$\sigma_{\theta_e^+}$  & \boldmath$\sigma_{E_\gamma}$ & \boldmath$\sigma_{x_\gamma}$ & \boldmath$\sigma_{t_{e^+\gamma}}$ & \boldmath$\epsilon_{e^+}$ & \boldmath$\epsilon_\gamma$\\
    \midrule
    MEG           & 380 keV/c  & 9.4 mrad  & 2.4\%/1.7\%  & 5 mm     & 122 ps   & 30\%   & 63\%\\
    MEG-II        & 100 keV/c  & 6.7 mrad  & 1.7\%/1.7\%  & 2.4 mm   &  70 ps   & 65\%   & 69\%\\
    \bottomrule
  \end{tabularx}
\end{table}

Recently, the MEG collaboration worked on the upgrade of the experiment (MEG II), which aims to reach a sensitivity of $6\times 10^{-14}$ \@ 90\% C.L.~\cite{meg_ii}. Various improvements on the detector were delivered. The positron spectrometer was replaced with a low-mass single-volume cylindrical drift chamber with high rate capability. This increased the acceptance of the spectrometer with respect to the MEG configuration by more than a factor of 2. The LXe calorimeter was also upgraded by replacing the MEG photomultiplier tubes (PMTs) with smaller vacuum-ultraviolet sensitive silicon photomultipliers (SiPMs). A novel timing detector for an active suppression of the accidental background was also introduced. The results of the engineering runs showed a fast degradation of the wires of the drift chamber and of the SiPMs~\cite{meg_ii}. Table~\ref{tab:meg_perf} compares the new detector performance with the previous ones reported for the MEG detector. The MEG-II collaboration plans to build a new chamber to replace the existing one, and they will take advantage of the coming engineering runs to study more carefully the degradation of the SiPMs. Preliminary results show that they can adjust the operation conditions to achieve the desired level of sensitivity~\cite{meg_ii}. 

\subsubsection{\mueee}
In the \mueee decay, the final state consists of two positrons and one
electron emerging from the same vertex with an invariant mass that
matches the muon rest mass. In a three-body
decay, the energy associated to each product is not a fixed
amount. Simple relativistic kinematics consideration show that the
maximum energy of one of the decay products is 
about $m_\mu /2$ 
and that the decay can be described by two independent variables. 
The energy distribution of each daughter particle depends on the exact dynamics of the underlying
unknown physics. In general, the highest energy particle is expected
to have a momentum larger than 35 MeV/c, while the distribution of the
lowest energy particle peaks near zero and decreases quickly as its
energy tends to its upper limit so that only about one half have an
energy larger than 15 MeV~\cite{Signorelli}.
The background sources for this search can be factorized in two main
categories: a physical background coming from the $\mu^+\to
e^+ \bar{\nu}_\mu \nu_e e^- e^+$ process, and an uncorrelated
component coming from the accidental coincidence of a positron from a
Michel decay and a positron-electron pair produced by the interactions
of other positrons or muon with the target or the detector
material. The accidental background component scales quadratically
with the muon beam intensity. As in the \meg case, it is more
convenient to design an experimental apparatus that uses positive
muons.

\begin{figure}[H]
    \includegraphics[width =0.7\textwidth]{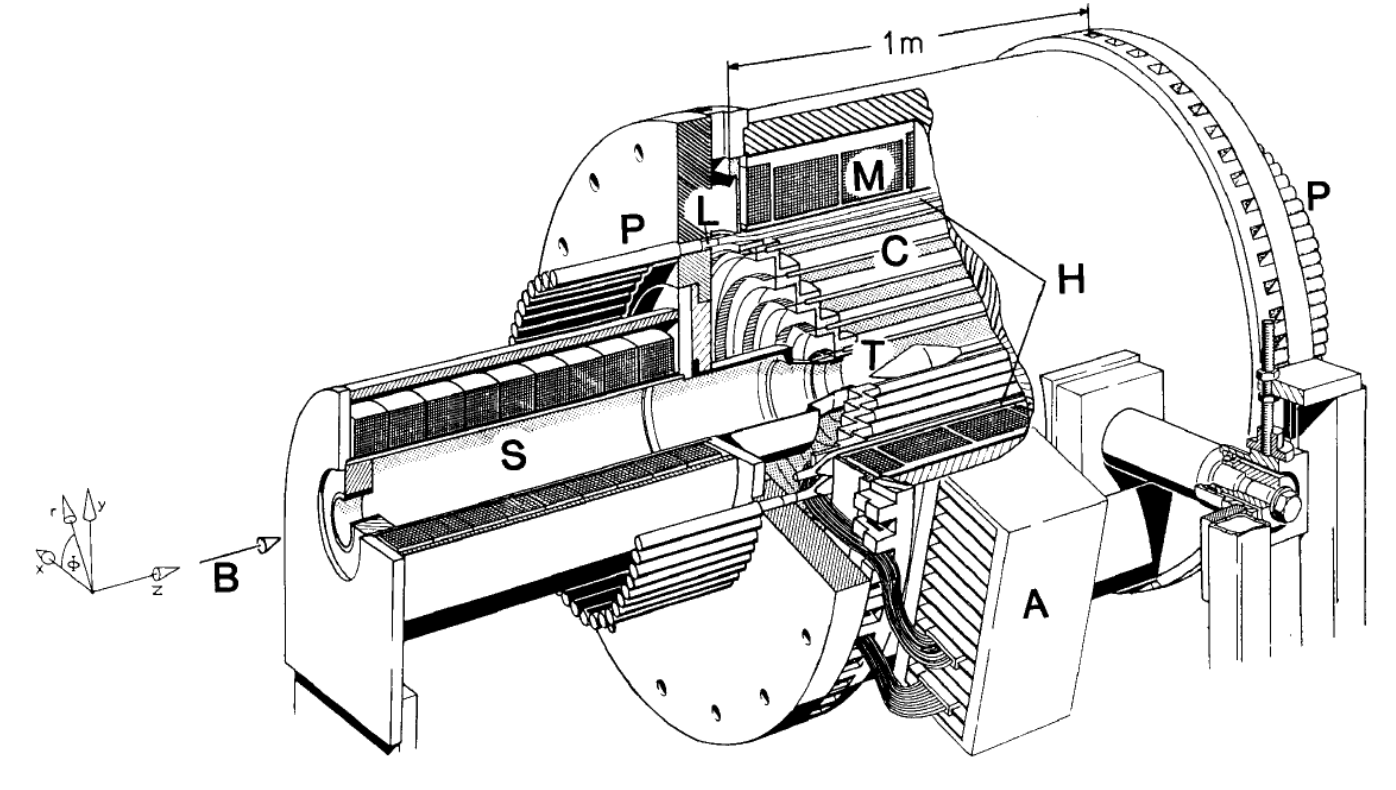}
    \caption{Schematic of the SINDRUM experiment.  B, muon beam; S, focussing solenoid; T, target; C, five cylindrical multi-wire proportional chambers; H, hodoscope of 64 scintillators; L, light guides for the hodoscope; P, 128 photomultipliers; A, preamplifiers for the cathode strips and amplifier/discriminators for the anode wires; M, normal conducting coil of the magnet. Figure and caption from~\cite{sindrum}.}\label{fig:sindrum}
\end{figure}

The current best limit on \mueee, $1.0\times 10^{-12}$~\cite{BELLGARDT19881}
at 90\% C.L., was set by the SINDRUM experiment at
PSI~\cite{BELLGARDT19881} based on $\sim$10$^6$ stopped  $\mu^+$. The SINDRUM
apparatus, shown in Figure~\ref{fig:sindrum}, consisted of a double
cone-shaped stopping target in
the middle of five concentric multi-wire proportional chambers
surrounded by an array of plastic scintillator counters inside a
solenoidal magnetic field. For a 50 MeV electron/positron, the
detector apparatus had a momentum resolution at the level of $\sim
$1~MeV/c, a timing \mbox{resolution $\leq 1$~ns} and a vertex resolution of
$\sim$1~cm. The data reduction was achieved with a multiple stage
trigger, taking advantage of track and charge pre-filters that were requiring 
at least one negatively and two positively charged tracks within a
time window of 7~ns. Then, a track-correlator was used to limit
 the total transverse momentum of the $e^+e^-e^+$ triplet 
below 17~MeV/c. In the statistical analysis, the event candidates were
determined from the two-dimensional distribution of $\sum E_i$ vs.
$\hat{p}^2$, where $\hat{p} = \left( p_L/\sigma_L\right)^2 + \left(
p_T/\sigma_T\right)^2$ ($L$ and $T$ denote the longitudinal and
transverse components with respect to the beam axis). This parametrization is
particularly convenient because the signal candidates satisfy $\sum E_i =
m_\mu$ and $\hat{p}^2$ is expected to peak near 0.

\begin{figure}[H]
    \includegraphics[width =0.7\textwidth]{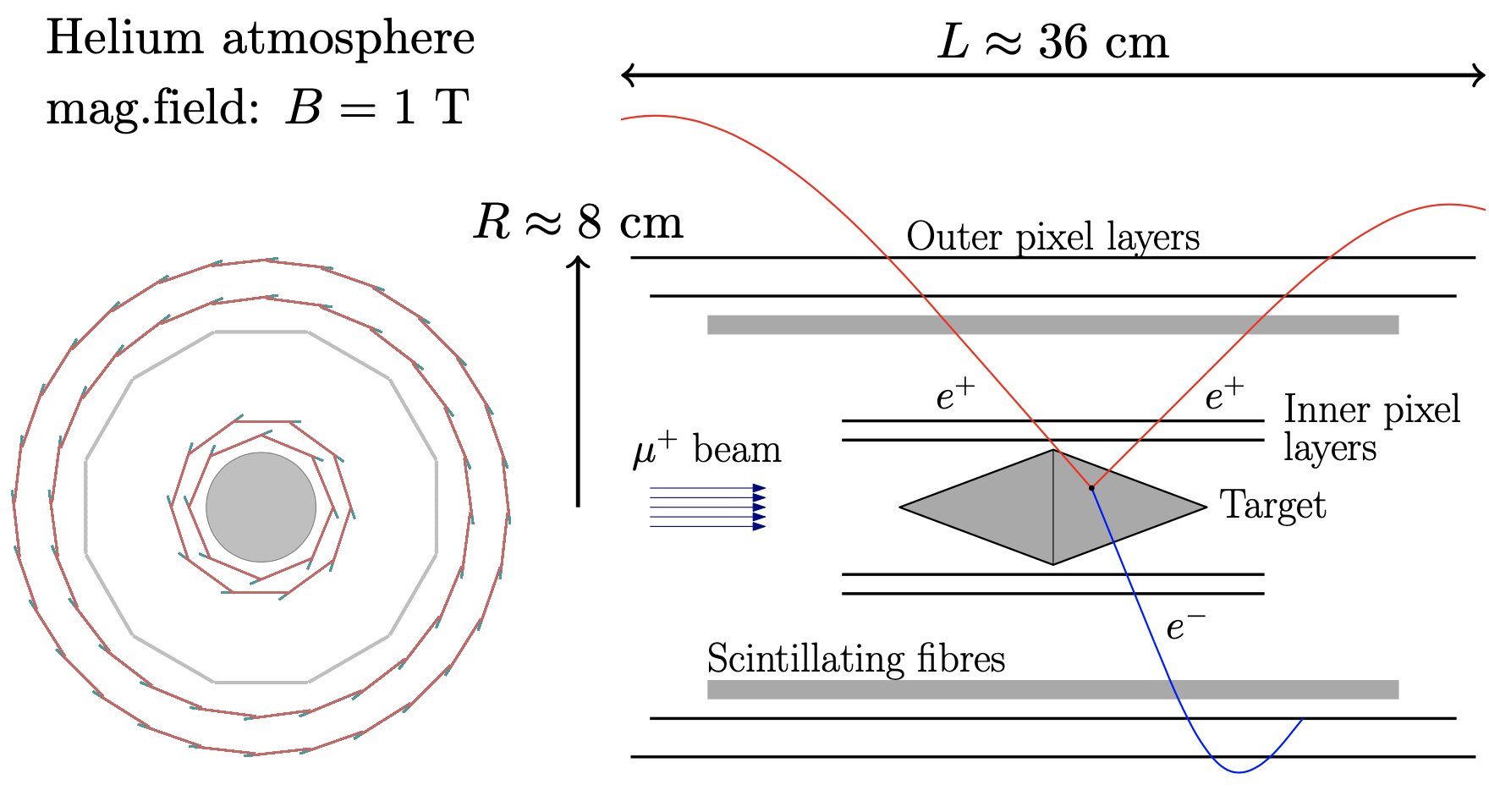}
    \caption{Schematic of the Mu3e experiment (Figure from~\cite{mu3e_2}).}\label{fig:Mu3e}
\end{figure}

A new effort to improve the sensitivity on \mueee search is
underway at PSI by the Mu3e collaboration~\cite{mu3e}. The Mu3e 
experiment aims for a $10^{-16}$ single-event sensitivity,
which would correspond to an improvement by four orders of magnitude
compared to the limit set by the SINDRUM experiment. Such a leap in
sensitivity is enabled by: (i) the availability of high-intensity muon
beams, (ii) the use of silicon pixel detectors instead of multi-wire
proportional chambers to track the decay products, and (iii) a modern
data-acquisition system able to handle the vast amount of data
produced by the detector. A first phase of the
experiment is currently under construction at the $\pi$E5 beamline at
PSI, where the intense DC surface muon beam of $10^8 \mu^+/s$ will be
exploited to achieve a single event sensitivity of $2\times10^{-15}$
in about 300 days of data taking~\cite{mu3e_2}. The Mu3e experimental
setup is shown in Figure~\ref{fig:Mu3e}. It is designed to track the
two positrons and one electron from the positive muon decaying at rest with a
light-weight tracker placed inside a 1 T magnetic field, thereby
reconstructing the decay vertex and invariant mass.
The muon beam is stopped in a hollow double-cone target placed 
at the center of the Mu3e solenoid. This allows for the spread out of 
the decay vertices in z and minimizes the amount of target material
traversed by the decay particles. The target is surrounded by the
cylindrical central tracker, which consists of an inner silicon pixel
detector, a scintillating fiber tracker for time measurements, and an 
outer silicon pixel detector. A momentum resolution of better than 1
MeV/c at @ 50 MeV/c is achieved by letting the positrons (electrons) re-curl in the
magnetic field, either crossing the central tracker again or hitting
the outer tracking stations surrounding the upstream and downstream
beam pipe. These stations consist of a silicon pixel tracker and a
scintillating tile detector mounted on the inside of the pixel
tracker. The 5~mm thick tiles enable a time resolution for the tracks
reaching these outer stations of better than 100~ps. The material
budget, which must be minimized to reduce the multiple scattering and thus
deliver the required momentum resolution, was minimized by means of
custom High-Voltage Monolithic Active Pixel Sensor (HV-MAPS)~\cite{mu3e_maps}
based on a commercial 180~nm HV-CMOS process. Together with
its support structure, the entire silicon tracking module has a
thickness of $\sim$0.12\% radiation lengths, with a single-hit
efficiency $>99\%$ and a time resolution of $O(10 \ \mathrm{ ns})$. A gaseous
helium cooling system allows the experiment to dissipate 250 mW/cm$^2$
of power generated by the MAPS modules.
A time resolution of about 10 ns is insufficient to determine the
direction and thus the charge of the decay particles. A scintillating
fiber detector is, therefore, placed between the inner and outer layer
of the central silicon-pixel tracker, consisting of a dozen 30 cm long
ribbons made from three staggered layers of 250~\textmu m diameter
multi-clad round fibers, read out by Silicon Photomultipliers (SiPM)
arrays on both sides. Located at the very end of the re-curling
particle trajectories hitting the upstream or downstream tracker,
where the constraints on the material budget are less stringent, the
tile detector provides the needed precise timing information of the
particle tracks, in conjunction with the fiber detector, significantly
reducing the accidental background. Each tile is read out by a single
SiPM. For the tile and fiber detector, a time resolution of $<$50 ps
and $<$400 ps is achieved, respectively.
Mu3e had a successful integration run campaign from May to July 2021
with a reduced detector: 2 pixel layers + fiber detector.

\subsubsection{\muconv}
\begin{figure}[H]
    \includegraphics[width =0.7\textwidth]{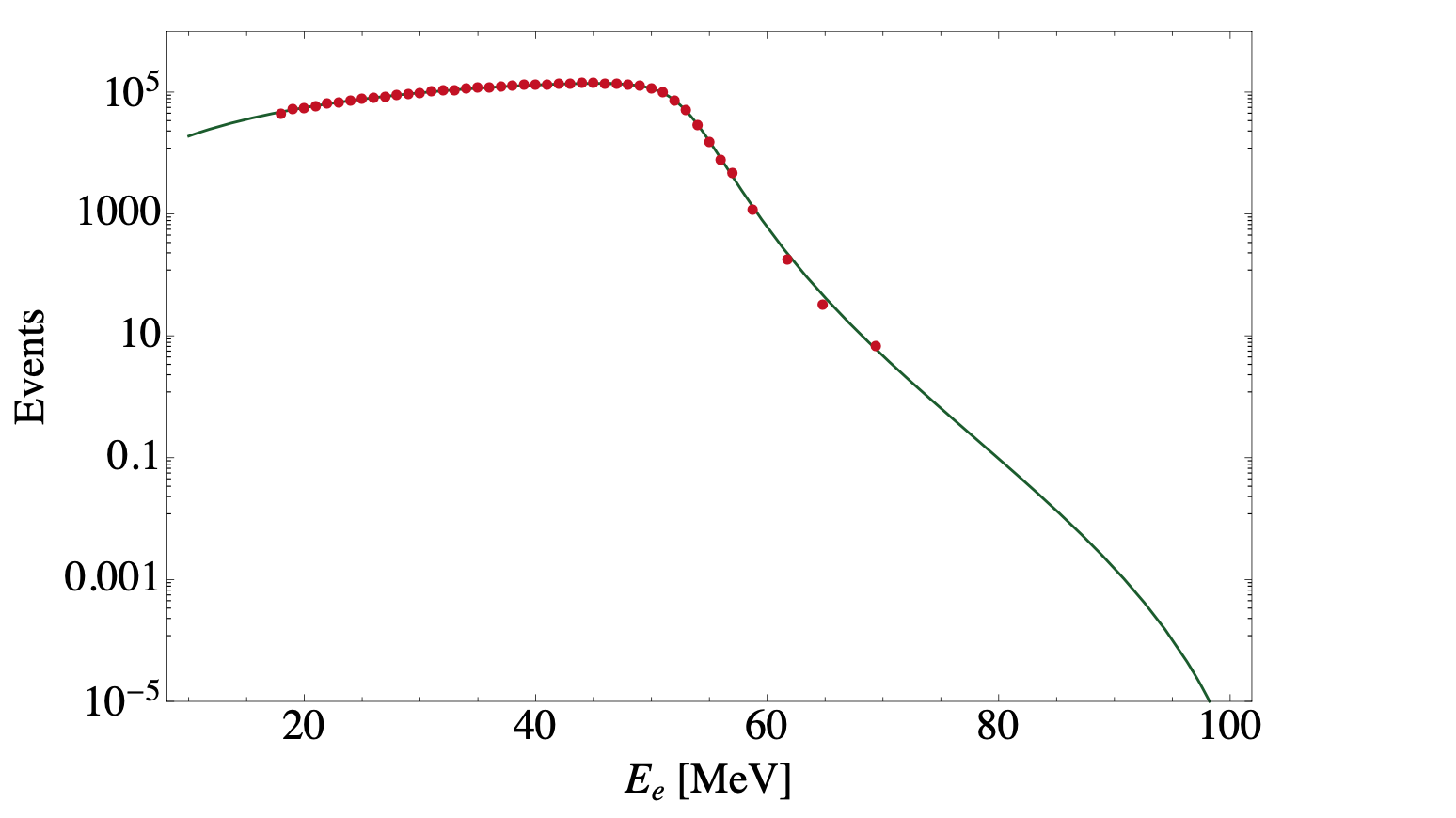}
    \caption{Energy spectrum of the  DIO electrons (solid line) fitted to TWIST data (dots
)~\cite{TWIST}. Figure from~\cite{DIOSpectrum}. }\label{fig:DIO}
\end{figure}

\muconv{} conversion is the process where a muon converts into an electron in
the field of a nucleus without producing neutrinos in the final
state. This process has the same dynamic of a two-body decay and, therefore, 
results with a monochromatic electron with an energy $E_{\mu e}$:
\begin{equation}
    E_{\mu e} = m_{\mu} - E_b - \frac{E_{\mu}^2}{2 m_N}\nonumber,
\end{equation}
where $m_{\mu}$ is the muon mass, $E_b\sim Z^2\alpha^2 m_{\mu}/2$ is the muonic binding energy and the last term is from nuclear recoil energy up to terms of order $1/m_{N}^2$, neglecting variations of the weak-interaction matrix element with energy~\cite{Bernstein}, where $E_\mu = m_\mu - E_b$ and $m_N$ is the atomic mass. In the case of Al, which is the selected material for the current experiments under construction, $E_{\mu e} \sim$ 104.96~MeV.  In muon conversion experiments, the quantity
\begin{equation}
    R_{\mu e} = \frac{\Gamma(\muconv)}{\Gamma(\mu^- + N  \to all-capture)}\nonumber
\end{equation}
is measured. The normalization to captures simplifies calculations as 
many details of the nuclear wavefunction cancel
in the ratio~\cite{MutoeKitano}.  The coherent conversion
leaves the nucleus intact, and there is only one detectable
particle in the final state. As we will see, the resulting electron energy stands
out from the background, hence muon-electron conversion does not suffer from
accidental background, and extremely high rates can be used.
Negative muons stopped in the stopping target can undergo a nuclear
capture. Particles generated in the muon capture (n, p and $\gamma$)
may reach the detector system and create extra activity that can
either obscure a conversion electron (CE) track or create spurious
hits. As a result, some specific shielding is required to reduce this
background. 
%
Electrons from the high momentum tail of the muon decay-in-orbit (DIO) represent the
intrinsic background source for the \muconv{}
search. Figure~\ref{fig:DIO} shows the energy spectrum of DIO
electrons~\cite{DIOSpectrum}.

The main features of the DIO energy spectrum can be summarized as follows:
\begin{itemize}
    \item 
    \textls[-25]{the endpoint of the spectrum corresponds to the energy of the electrons from \muconv{} conversion (CE);}
    \item 
    the overall spectrum is falling as $(E_{\mu } - E)^5$, where $E$ is the DIO energy;
    \item
    about $10^{-17}$ of the spectrum is within the last MeV from the endpoint.
\end{itemize}

Therefore, to reach a high sensitivity at the level $O(10^{-17})$, the
detector resolution is crucial. 
As the muon beam is generated from charged pions, another relevant
background comes from the radiative pion capture (RPC) process $\pi^-
N \to \gamma N^*$, followed by the electron-positron pair conversion
of the photon. Unfortunately, not all pions decay in the transport
line, and, consequently, the muon beam is contaminated by pions. This
source of background is reduced by taking advantage of the difference
between the pion and the muonic atom lifetimes. The pion has a decay
constant $\tau$ < few tens of ns, while the bound muon has a $\tau$ of
the order of several hundreds of ns (depending on the Z of the
material). Therefore, using a pulsed beam structure, it is possible to
set a live gate delayed with respect to the beam arrival, reducing 
the RPC contribution to the desired level. Other
beam-related sources of background are: remnant electrons in the beam
that scatter in the stopping target, muon decays in flight and
antiprotons interacting in the apparatus.
Atmospheric muons can also represent a significant source of background
because these particles can interact in the apparatus and eventually
generate a signature very similar to the CE. An active shielding is thus required to detect the incoming cosmic muons crossing the apparatus and veto the event candidates on time. Moreover, the detector system 
has to provide particle identification (PID) capabilities to reject 
un-vetoed muons that can mimic the CE due to a mis-reconstruction.

\begin{figure}[H]
    \includegraphics[width =0.7\textwidth]{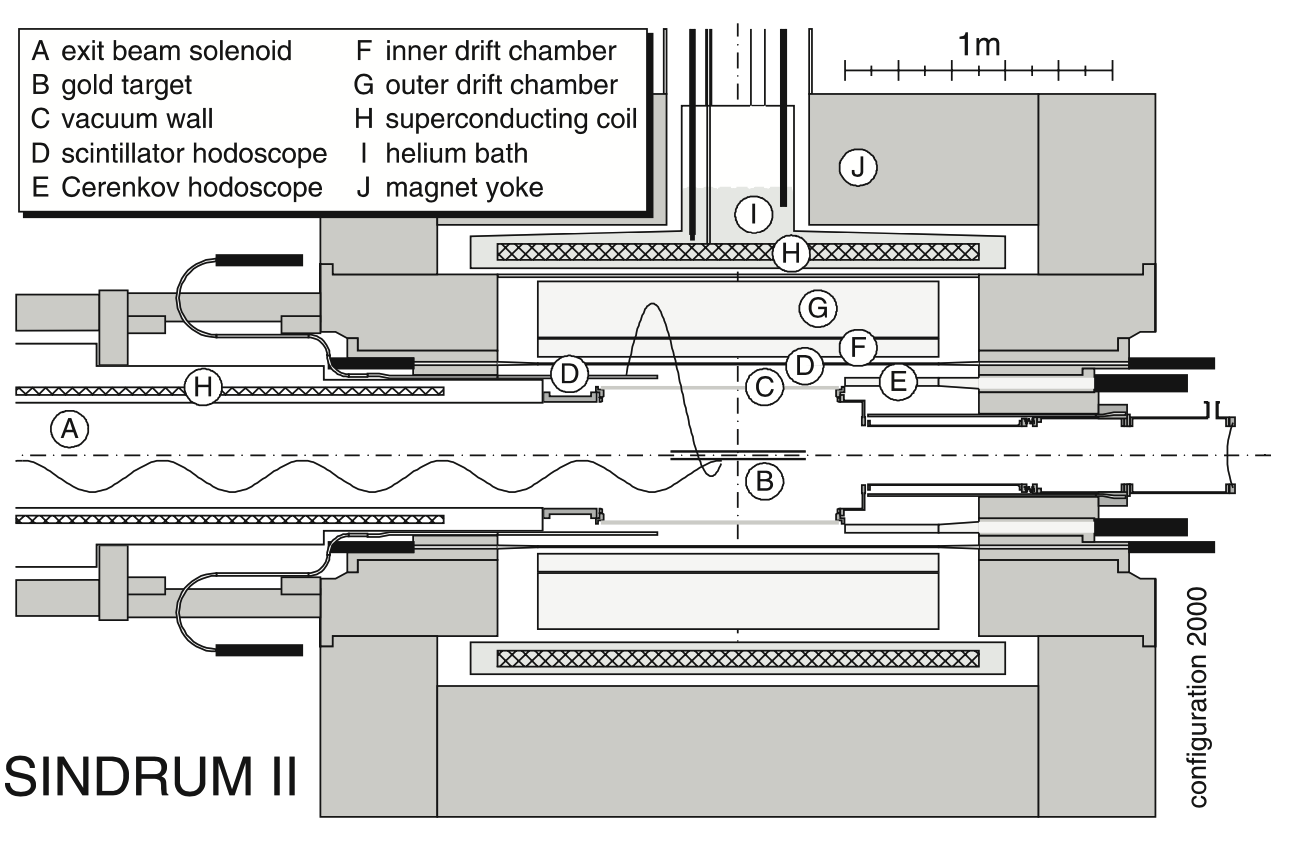}
    \caption{The SINDRUM-II experimental setup. Figure from~\cite{SINDRUM_II}.}\label{fig:sindrumII}
\end{figure}

The current best limit on the \muconv{} measurement comes from the
SINDRUM-II experiment at PSI~\cite{SINDRUM_II}. In SINDRUM-II, a high
intensity muon beam was stopped in a target that was surrounded by the
detector elements housed in a superconducting
solenoid. Figure~\ref{fig:sindrumII} shows a sketch of the SINDRUM-II
apparatus. The detector consisted of two drift chambers, to
reconstruct the trajectories of the charged particles, and Cerenkov
hodoscopes, to measure the timing of the reconstructed tracks and for
providing PID capabilities.
With a total of $\sim$10$^{14}$ stopped muons, SINDRUM-II reached a sensitivity at the level of $\sim$10$^{-13}$ on the  \muconv{} process using different target materials~\cite{SINDRUM_II}.

New experimental concepts have been proposed
and are currently under construction at Fermilab (USA) and J-PARC 
(Japan) to search for \muconv{} with unprecedented sensitivity at
the level of $\sim$10$^{-17}$.
The Mu2e experiment at Fermilab had its genesis back in the 1980s,
behind the Iron Curtain. In a way, Mu2e was born in the Soviet
Union. In 1989, the Soviet Journal of Nuclear Physics published a
letter to the editor from physicists Vladimir Lobashev and Rashid
Djilkibaev, where they proposed an experiment that would perform the
most thorough search yet for muon-to-electron flavor violation. In
1992, they proposed the MELC experiment at the Moscow Meson
Factory~\cite{MELC}, but in 1995, due to the political and economic
crisis, the experiment shut down. The same overall scheme was
subsequently adopted in the Brookhaven National Laboratory MECO
proposal in 1997~\cite{MECO} and then in the Mu2e and COMET experiments. 

\begin{figure}[H]
    \includegraphics[width =0.7\textwidth]{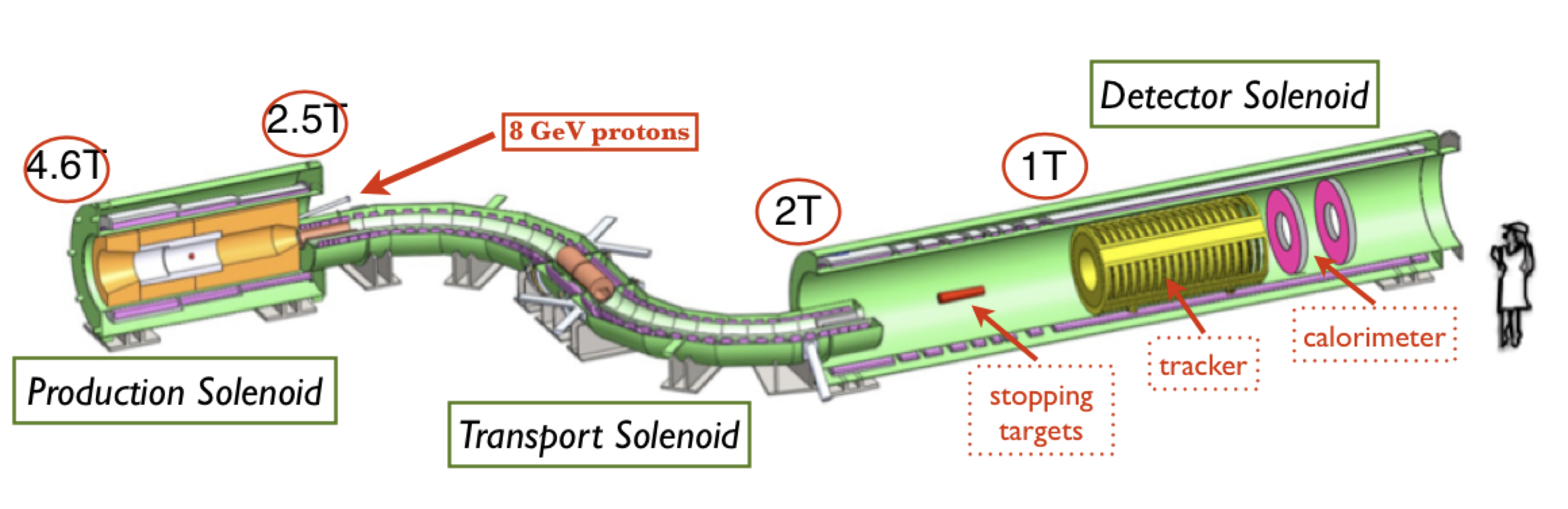}
    \caption{Schematic of the Mu2e experiment.}\label{fig:Mu2e}
\end{figure}

The Mu2e apparatus~\cite{mu2e}, shown in Figure~\ref{fig:Mu2e},
consists of three main superconducting solenoids. The first two, named
production and transport solenoid in Figure~\ref{fig:Mu2e}, are used to generate a
high-intensity, low-momentum muon beam starting from a 8 GeV proton
beam. The third solenoid, named $"$Detector Solenoid$"$ in
Figure~\ref{fig:Mu2e}, contains an Al stopping target, where the muons
are stopped to generate the muonic atoms, and downstream to it, we have
a low-mass straw-tube tracker~\cite{mu2e:tracker}, followed by a pure-CsI
crystal calorimeter~\cite{mu2e:calorimeter}. Both detectors are left
un-instrumented in the inner 38 cm to avoid any interaction with the
largest majority ($>$99\%) of the low momenta electrons coming from
the muon DIO processes in the stopping target. In Mu2e, the stopping
target is not placed in the middle of the tracker as it was done in
SINDRUM-II to limit the flux of protons, photons and neutrons (from
the muon nuclear captures) in the detector. A graded magnetic field
around the stopping target increases the detector geometrical
acceptance by reflecting the electrons that initially were emitted in
the direction opposite to the detector. The whole detector solenoid
and half of the transport solenoid are covered with a cosmic ray veto
system designed to detect atmospheric muons with an efficiency
$\geq$99.99\%.

\begin{figure}[H]
    \includegraphics[width =0.7\textwidth]{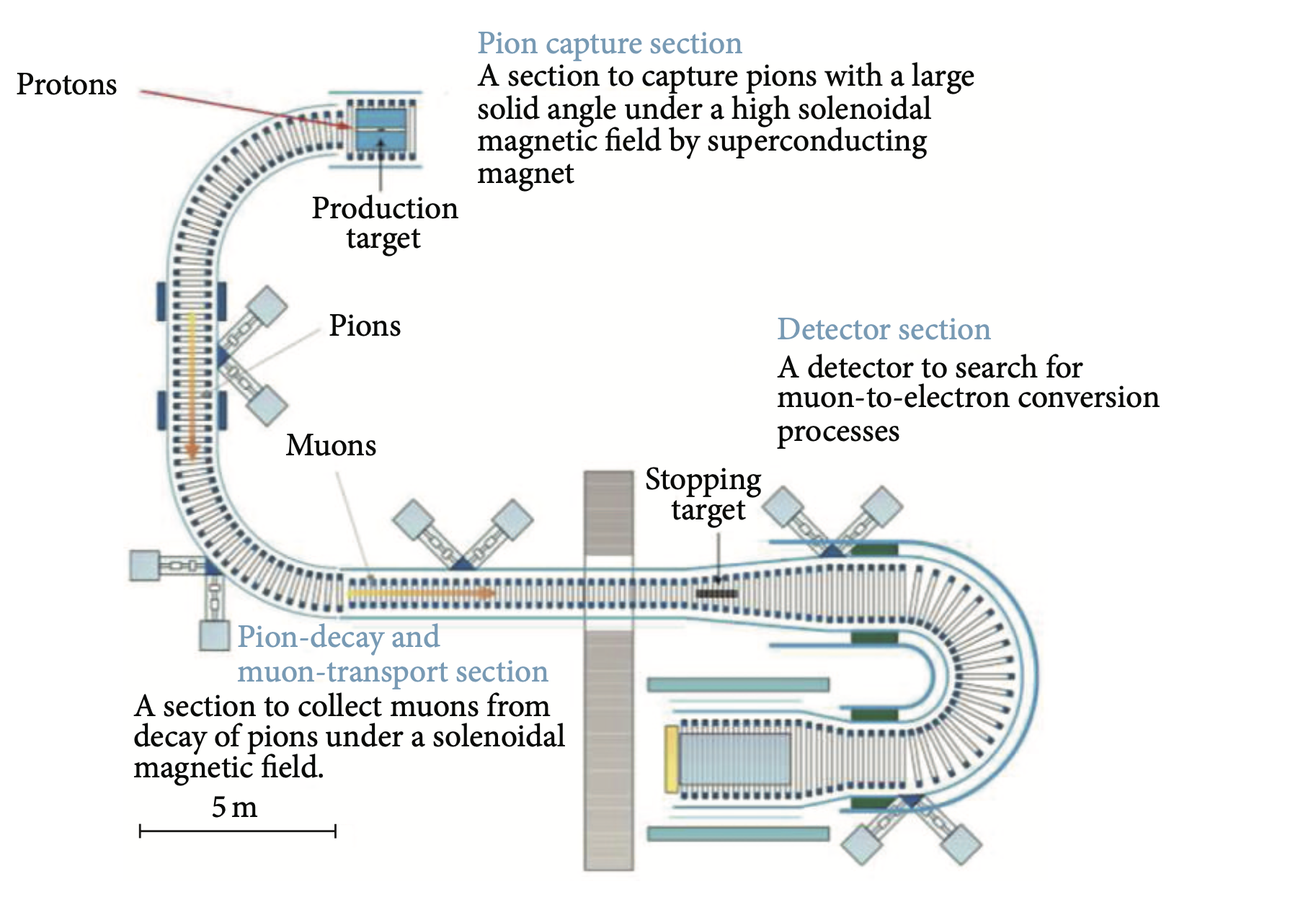}
    \caption{Schematic layout of COMET Phase-II (Figure from
~\cite{COMET_PHASE_I}).}\label{fig:COMET}
\end{figure}

The design of the COMET experiment at J-PARC, shown in
Figure~\ref{fig:COMET}, is based on a similar concept. 
A 8 GeV pulsed proton beam is used to produce pions, which are
then captured and transported by a series of superconducting
solenoids. The pions decay into muons as they travel along the muon
transport channel. The toroidal field of the muon transport channel
selects muons with negative charge and momentum less than 75
MeV/c. The major difference with respect to the Mu2e design is that a
second transport line is installed between the muon stopping target
and the detector regions to select charged particles of momentum
centered around 100~MeV/c. The detector system consists of a straw-tube
tracker followed by a LYSO crystal calorimeter~\cite{COMET_PHASE_I}.

COMET plans to operate in two stages: Phase-I and Phase-II. Phase-I will allow the experiment to characterize the beam and the key backgrounds as
well as provide enough statistic to reach a 90\% C.L. sensitivity at
the level of $7\times10^{-15}$~\cite{COMET_PHASE_I}. During Phase-I,
COMET will operate with a smaller apparatus that consists of half of
the first C-shaped muon transport line directly connected to a
solenoid that houses the muon stopping target surrounded by the
detector system. For Phase-I, the detector consists of a cylindrical
drift chamber and a set of trigger hodoscope counters.

Another experiment, named DeeMe~\cite{deeme}, aims to search for
the \muconv{} process with a single event sensitivity of
$1\times10^{-13}$ using a graphite target. The experiment is conducted
at the Materials and Life Science Experimental Facility (MLF) of the J-PARC. Muonic atoms are produced in a primary-proton target itself, which is hit by pulsed proton beams from the Rapid Cycling Synchrotron (RCS) of J-PARC. To detect the electron and measure its momentum, a magnetic
spectrometer, consisting of a dipole magnet and four sets of multi-wire
proportional chambers (MWPCs)~\cite{deeme_mwpc}, is employed. 
The spectrometer is expected to reach a resolution of 
$\sigma_p < 0.5$ MeV/c at 100~MeV/c. 
The resolution is needed to reject the DIO background, which is
the dominant source of high energy electrons for this search.
The number of charged particles hitting the detectors is estimated 
with Monte Carlo simulation to be approximately $10^8$ particles 
per proton-bunch with an RCS power of 1 MW.
The construction of the secondary beamline for DeeMe, the H Line, is now
in progress. Meanwhile, the collaboration measured the momentum
spectrum of the DIO electrons in the momentum region 48--62~MeV/c 
at the D2 area at MLF. This measurement will be important for validating the theoretical models used to model the DIO background and characterize the detector performance. Three sets of measurements were performed between the year
2017 and 2019~\cite{deeme}, and the analysis is now underway.

\subsubsection{\muconvp}
\muconvp{} conversion is the process where a muon converts into an positron in the field of a nucleus that undergoes a nuclear transition. This process violates the lepton number ($\Delta L=2$) and the lepton flavor conservation. The experiments looking for the \muconv{} process can typically search for the \muconvp{} as well. The current best limit on the \muconvp{} process comes from the SINDRUM-II experiment~\cite{1998334} that set a limit at $5.7 \times 10^{-13}$ at 90\% C.L. The major background source is the radiative muon capture, where the photon can generate (via asymmetric conversion) a positron with an energy close to the signal region. 

The search for the \muconvp{} complements the $0\nu\beta\beta$ decay searches and is sensitive to potential flavor effects in the neutrino mass-generation mechanism.
We refer the reader to~\cite{muconvp} for additional information about the current status and future prospects offered by the COMET and Mu2e experiments.

\subsection{CLFV Searches Using Taus}\label{sec:tau_sector}
\begin{figure}[H]
     \includegraphics[width =0.79\textwidth]{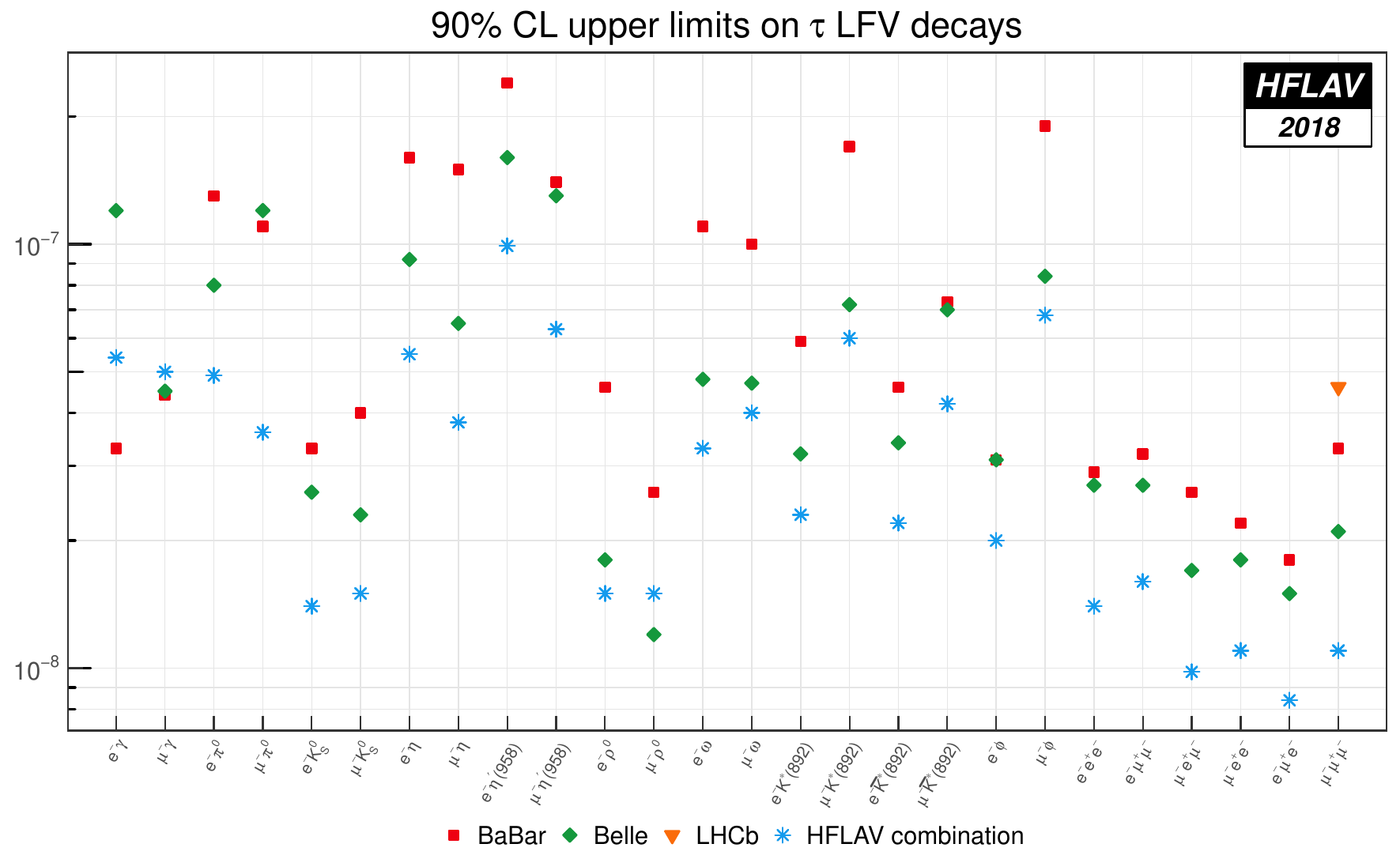}
     \caption{Tau lepton-flavor-violating branching fraction upper limits 
     combinations summary plot. For each channel, we report the HFLAV 
     combined limit and the experimental published limits. In some cases, 
     the combined limit is weaker than the limit published by a single experiment. 
     This arises since the CLs method used in the combination can be more 
     conservative compared to other legitimate methods, especially when 
     the number of observed events fluctuates below the expected  \linebreak
     background~\cite{HLFV2018}.}\label{fig:Tau_limits}
\end{figure}

The tau lepton is, in principle, a very promising source of CLFV decays. 
Thanks to the large tau mass ($m_\tau \approx 1.777$ GeV), many CLFV 
channels can be investigated: \taumug, \taueg, \taulll, \taulh, ... 
($l=e,\ \mu$ and $h$ is a light hadron).
Table~\ref{tab:clfv_limits} lists the current best limits on the 
tau CLFV searches, and Figure~\ref{fig:Tau_limits} shows the individual 
results from the BaBar~\cite{BABAR}, Belle~\cite{BELLE} and the 
LHCb~\cite{LHCB} experiments, together with their combination.

From the experimental point of view, however, a difficulty immediately 
arises: the tau is an unstable particle, with a very short lifetime 
($\tau = 2.91 \times 10^{-13}$ s). As a result, tau beams cannot be realized, 
and large tau samples must be obtained in intense electron or proton 
accelerators, operating in an energy range where the tau production 
cross section is large.  
At $e^+e^-$ and $pp$ collider machines, the majority of the tau particles are not produced at rest, which means that, unlike the muon searches discussed before, here we need to deal with decays-in-flight. Thanks to the boost, the decay products could get energy values up to several GeV, which experimentally poses the challenge to deliver wide-range calibrations for the detectors (from a few hundreds of MeV to several GeV). 
For all these searches, events contain 
 a pair of taus in which one tau undergoes SM decay (tag side), while the signal side is selected on the basis of the appropriate topology of each individual channel. The tagging side accepts the leptonic ($\tau\to l \nu \bar{\nu}$) and 1-prong hadronic  decays, while on the signal side, CLFV candidates are selected on the basis of the appropriate topology of each individual channel.
The following paragraphs discuss the current best limits for some of these experimental searches from experiments at $B$-factories and $pp$ colliders.

\subsubsection{$\tau \to l \gamma$}
\begin{figure}[H]
 \includegraphics[width
    =0.8\textwidth]{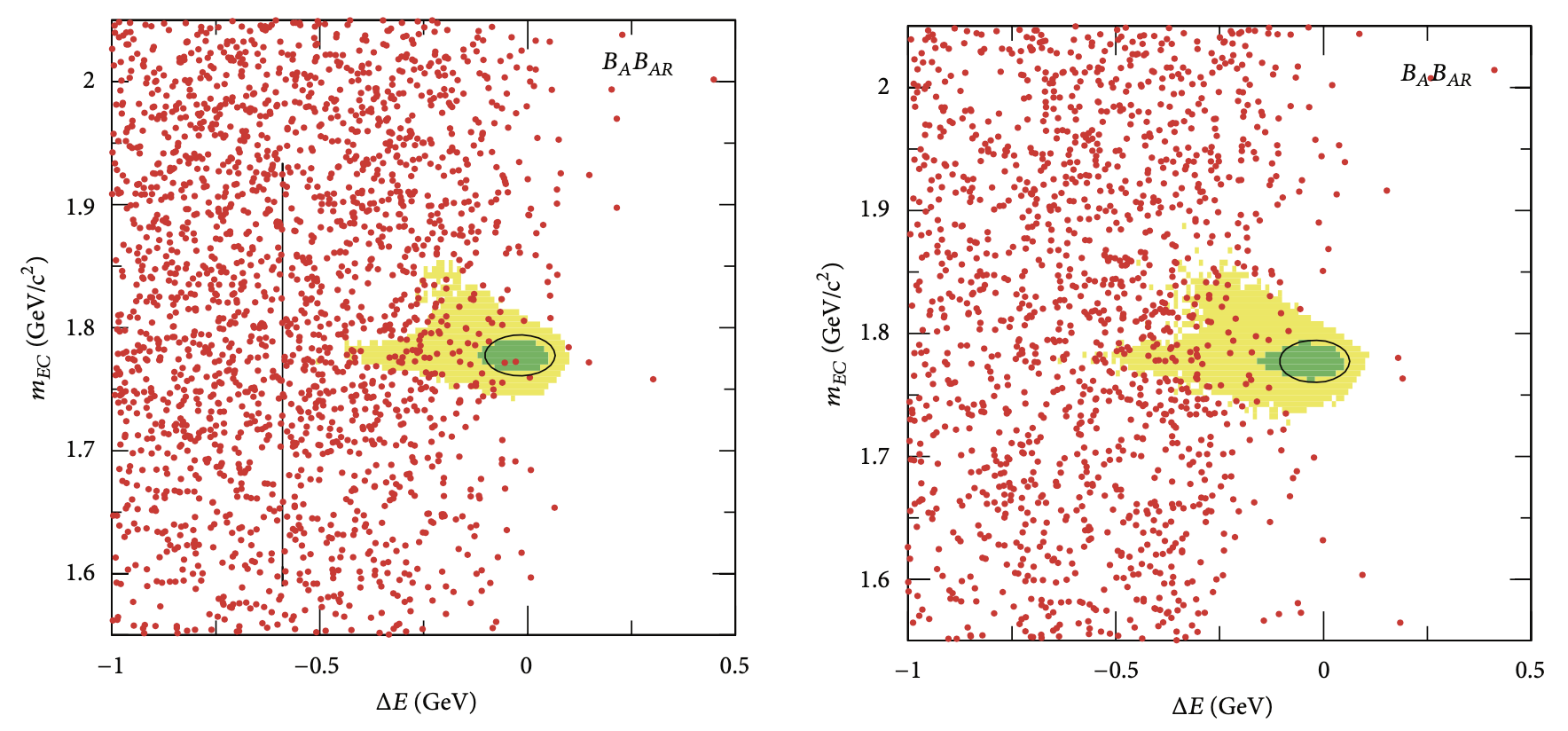} \caption{The
    Grand Signal Box and the $2\sigma$ ellipse for $\tau^\pm \to
    e^\pm \gamma$ (\textbf{left}) and $\tau^\pm \to \mu^\pm \gamma$ (\textbf{right})
    decays in the $m_{BC}$ vs. $\Delta E$ plane. Data are shown as dots,
    and contours containing 90\% (50\%) of signal MC events are shown
    as light-shaded (dark-shaded) regions (Figure and caption
    from 
~\cite{babar_taulg}).}\label{fig:babartaulg}
\end{figure}

The $\tau \to l \gamma$ decay, where $l$ is a light lepton ($e,\ \mu$),
has been one of the most popular CLFV tau channels. The signal is
characterized by a $l^\pm - \gamma$ pair with an invariant mass and
total energy in the center-of-mass (CM) frame ($\mathrm{E_{CM}}$)
close to $m_\tau = 1.777$ GeV and $\sqrt{s}/2$, respectively. 
The dominant irreducible background comes from $\tau$-pair events
containing hard photon radiation and one of the $\tau$ leptons
decaying to a charged lepton. The remaining backgrounds arise from the
relevant radiative processes, $e^+e^- \to e^+e^- \gamma$ and
$e^+e^- \to \mu^+\mu^- \gamma$ and from hadronic $\tau$ decays where a
pion is misidentified as an electron or muon.
For this decay channel, the current best limits comes from the BaBar
and the Belle collaborations. BaBar collected $(963 \pm 7) \times
10^6\ \tau$ decays near the $\Upsilon(4S),\ \Upsilon(3S)$ and
$\Upsilon(2S)$ resonances. In the BaBar detector~\cite{BABAR}, charged
particles are reconstructed as tracks with a 5-layer silicon vertex
tracker and a 40-layer drift chamber inside a 1.5 T solenoidal
magnet. A CsI(Tl) electromagnetic calorimeter is used to identify
electrons and photons. A ring-imaging Cherenkov detector is used to
identify charged pions and kaons. The flux return of the solenoid,
instrumented with resistive plate chambers and limited streamer tubes,
is used to identify muons. Signal decays are identified by two 
kinematic variables: the energy difference 
$\Delta E = E_{CM} - \sqrt{s}/2$ and the beam energy
constrained $\tau$ mass obtained from a kinematic fit after requiring
the CM $\tau$ energy to be $\sqrt{s}/2$ and after assigning the origin
of the $\gamma$ candidate to the point of closest approach of the
signal lepton track to the $e^+e^-$ collision
axis ($m_{BC}$). Figure~\ref{fig:babartaulg} shows the distributions of the
events for the two decay channels in $m_{BC}$ vs. $\Delta E$. The red
dots are experimental points, the black ellipses are the $2\sigma$
signal contours and the yellow and green regions contain 90\% and 50\% of MC
signal~events.
The searches yield no evidence of signals, and the experiment set upper limits on
the branching fractions of $B(\tau^\pm \to e^\pm \gamma) < 3.3\times
10^{-8}$ and $B(\tau^\pm \to \mu^\pm \gamma) < 4.4 \times 10^{-8}$ at
90\% confidence level~\cite{babar_taulg}.

\begin{figure}[H]
    \includegraphics[width =0.7\textwidth]{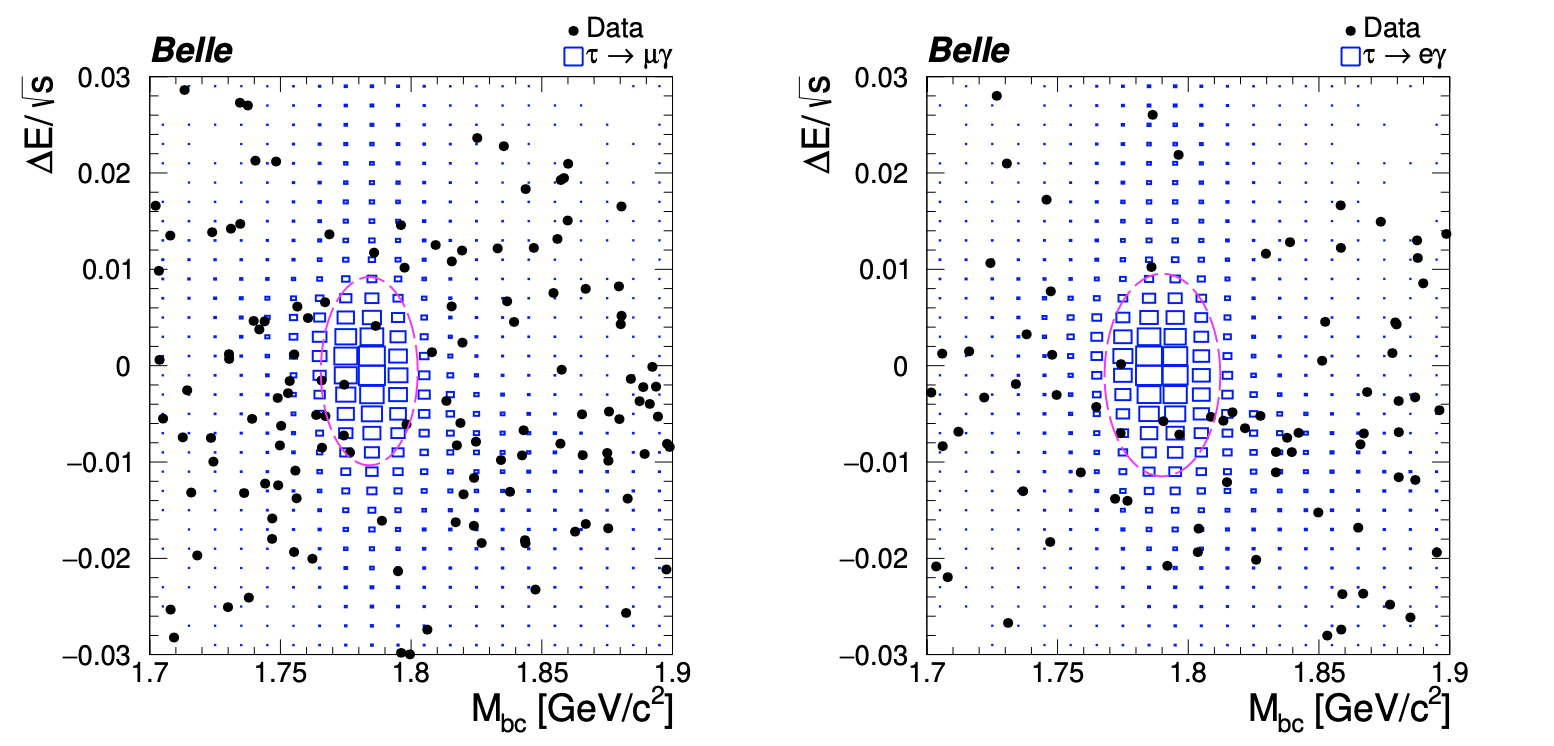}

    \caption{Two-dimensional distributions of $\Delta E/\sqrt{s}$ vs.
    $\mathrm{M_{BC}}$ for $\tau^\pm \to \mu^\pm \gamma$ (\textbf{left})
    and $\tau^\pm \to e^\pm \gamma$) (\textbf{right}) events. Black points are
    data, blue squares are $\tau^\pm \to l^\pm \gamma$ signal MC
    events, and magenta ellipses show the $\pm2\sigma$ signal regions used in this
    analysis (Figure and caption
    from 
~\cite{belle2021}).}\label{fig:belletaulg}
\end{figure}

The Belle experiment~\cite{BELLE} reported comparable limits using a
data analysis based on $988\ \mathrm{fb}^{-1}$ and a strategy similar to that of BaBar. Kinematical selections on missing momentum and opening
angle between particles are used to clean the
sample. Figure~\ref{fig:belletaulg} shows the two-dimensional
distribution of$\Delta E/\sqrt{s}$ vs. $m_{BC}$. The signal events have
$m_{BC} \sim m_\tau$ and $\Delta E/\sqrt{s} \sim 0$. The most dominant
background in the \taumug (\taueg) search arises from $\tau^+\tau^-$
events decaying to $\tau^\pm \to \mu^\pm \nu_\mu \nu_\tau$
($\tau^\pm \to e^\pm \nu_e \nu_\tau$) with a photon coming from
initial-state radiation or beam background. The $\mu^+\mu^-\gamma$ and
$e^+e^-\gamma$ events are subdominant, with their contributions
falling below 5\%. Other backgrounds such as two-photon and $e^+e^-\to q\bar{q} \ (q=u,\ d,\ s,\ c)$ are negligible in the signal region.

No significant excess over background predictions from the Standard
Model is observed, and the 90\% C.L. upper limits on the branching fractions are set at 
$B(\tau^\pm \to \mu^\pm \gamma) \leq 4.2 \times 10^{-8}$ and
$B(\tau^\pm \to e^\pm \gamma) \leq 5.6 \times 10^{-8}$~\cite{belle2021}. With the full dataset expected for the Belle II experiment~\cite{belle2_taulg} (the upgrade of Belle), 50 ab$^{-1}$, the upper limit for the branching fraction of LFV decays $\tau$ will be reduced by two orders of magnitude.

\subsubsection{$\tau \to 3 l$}

The signature for \taulll ($l=e,\ \mu$) is a set of
three charged particles, each identified as either an $e$ or a $\mu$,
with an invariant mass and energy equal to that of the parent $\tau$
lepton.

In the BaBar~\cite{babar_taulll} and Belle~\cite{belle_taulll}
analyses, all the six different combinations were explored. Events are
preselected requiring four reconstructed tracks and zero net charge,
selecting only tracks pointing toward a common region consistent with
$\tau^+\tau^-$ production and decay. The polar angles of all four
tracks in the laboratory frame are required to be within the
calorimeter acceptance range, to ensure good particle
identification. The search strategy consists of forming all possible
triplets of charged leptons with the required total charge and of
looking at the distribution of events in the ($m_{BC}$ , $\Delta E$)
plane ($m_{BC}$ and $\Delta E$ are defined as in the previous
section).
The backgrounds contaminating the sample can be divided in three broad
categories: low multiplicity $e^+e^-\to q\bar{q} \ (q=u,\ d,\ s,\ c)$ 
events, QED events (Bhabha or $\mu^+\mu^-$ depending on the specific 
channel) and SM $\tau^+\tau^-$ events. These background classes have distinctive distributions in the ($m_{BC}$, $\Delta E$) plane. 
The $e^+e^-\to q\bar{q} \ (q=u,\ d,\ s,\ c)$ events tend to
populate the plane uniformly, while QED backgrounds fall in a narrow
band at positive values of $\Delta E$, and $\tau^+\tau^-$ backgrounds
are restricted to negative values of both $\Delta E$ and $m_{BC}$ due
to the presence of at least one undetected
neutrino. Figure~\ref{fig:Tau_limits} shows the resulting limit for
all the combinations to be at the level of a few $10^{-8}$ for both
collaborations.

Even if the results are not yet competitive to those from 
$B$-factories, it is interesting to note that
experiments at the LHC have also been looking for the $\tau\to 3\mu$ decay.
The ATLAS experiment~\cite{atlas_taulll} performed a search for the
neutrinoless decay $\tau^- \to \mu^-\mu^+\mu^-$ using a sample of
$W^{-} \to \tau^-\bar{\nu}_\tau$ decays from a dataset corresponding
to an integrated luminosity of 20.3 fb$^{-1}$ collected in 2012 at a
center-of-mass energy of 8 TeV. The LHCb experiment~\cite{LHCb:2014kws}
performed the same search using a sample of tau from b and c-hadron
decays from a dataset corresponding to an integrated luminosity of 3.0
fb$^{-1}$ collected by the LHCb detector in 2011 and 2012 at
center-of-mass energies of 7 and 8 TeV, respectively. The CMS
experiment~\cite{cms_3l} recently delivered the results for the same
search using a sample of $\tau$ leptons produced in both W boson and
heavy-flavour hadron decays from a dataset corresponding to an
integrated luminosity of 33.2~fb$^{-1}$ recorded by the CMS experiment in 2016~\cite{cms_3l}.
ATLAS, CMS and LHCb reported a 90\% C.L. upper limit on the branching
ratio of $3.76 \times 10^{-7}$, $8.0 \times 10^{-8}$ and $4.6 \times
10^{-8}$, respectively.
The Belle-II collaboration studied prospects for the expected sensitivity
of this search. This channel has a purely leptonic final state, thus it 
is expected to be free of background. This allows to scale the experimental 
uncertainties linearly with the luminosity, which means that at least an 
improvement of a factor $\times 50$ is expected for Belle-II after accumulating 
a luminosity of 50 ab$^{-1}$~\cite{belle_ii}.

\section{Conclusions}
This review intended to provide a general theoretical and
experimental overview of CLFV processes.
CLFV processes are predicted by a large spectrum of
BSM theories. Limits set by the experiments are powerful to rule out
or heavily constrain the parameters space of many of these
models. In the theory sections, we discussed the CLFV phenomenology of several heavy physics scenarios. 
We reviewed CLFV signature of BSM theories that account for neutrino masses, discussing tree-level seesaw and models that generate neutrino masses radiatively, such as the scotogenic and Zee-Babu models.
Furthermore, we studied CLFV in the 2HDM and supersymmetric SM, and we presented a bottom-up analysis of LFV in effective field theories.

We also highlighted the state-of-the-art experiments involved in
direct CLFV searches in the muon sector. Aiming to measure branching ratios below $<$10$^{-11}$ requires very careful optimization  of the apparatus to limit the background contamination.  
The most recent results from the CLFV searches involving taus
performed by experiments at the LHC and $B$-factories were also
included.

In the coming decade, we expect to see many results delivered by
new muon CLFV direct searches (MEG-II, Mu3e, Mu2e, COMET) and by other experiments at the LHC and $B$-factories that could potentially improve the sensitivity to levels where many BSM theories expect a signal. This is possible thanks to formidable improvements in the muon beamline, novel detector and accelerator  technologies developed to face various experimental~challenges.

\vspace{6pt}
\funding{M.A. is supported by a doctoral fellowship from the Institut national de physique nucléaire et de physique des particules. G.P. is supported by Yale University with a GRANT from the US Department of Energy.}

\institutionalreview{Not applicable.}

\informedconsent{Not applicable.}

\dataavailability{Not applicable.}

\acknowledgments{M.A. thanks Sacha Davidson for useful discussions.}

\conflictsofinterest{The authors declare no conflict of interest.}

\appendixtitles{yes}

\appendixstart
\appendix
\section[\appendixname~\thesection]{\boldmath$ \mu \to e$ Conversion in Nuclei}\label{appendix:mutoeconv}
A muon, when stopped in a material, can form a muonic atom with a nucleus of the target. While in a bound state, the muon can undergo two SM processes: decay in orbit, where an electron and an (anti-)neutrino are emitted, or muon capture, given by
\begin{equation}
	\mu^-\ N(A,Z)\to \nu_{\mu}\ N'(A,Z-1)
\end{equation}
where $A,Z$ are, respectively, the mass and atomic number of the nucleus $N$. In the presence of LFV interactions that change muons to electrons, a muon can be captured by the nucleus without the emission of a neutrino
\begin{equation}
	\mu^-\ N(A,Z)\to e^-\ N(A,Z).
\end{equation}
in a processes known as $\mu\to e$ conversion in nuclei. 
After cascading down in energy levels, the ground state of the muonic atom is a 1s orbital with a binding energy $E_b$, and in the final state a monochromatic electron with energy $\sim m_\mu-E_b$   is emitted while the nucleus recoils. 
The SINDRUMII collaboration sets the upper limit $\Gamma(\mu N\to e N)/\Gamma_{capt}<7\times 10^{-13}$ \cite{SINDRUM_II} on the rate of $\mu\to e$ conversion with respect to the flavour conserving muon capture.

The state-of-the-art calculations for the conversion rate can be found in \cite{Kuno:1999jp, MutoeKitano}. In their notation, we describe coherent and spin-independent $\mu\to e$ conversion with LFV contact interactions among leptons and light quark currents  
\begin{align*}
	-\mathcal{L}_{conv}=2\sqrt{2}G_Fm_\mu( A_L \bar{e} \sigma^{\alpha\beta} P_L \mu F_{\alpha\beta}+ A_R \bar{e} \sigma^{\alpha\beta} P_R \mu F_{\alpha\beta}) \nonumber\\
	+\frac{G_F}{\sqrt{2}}\sum_{q=u,d,s}\bigg[(g_{LS(q)}\bar{e}P_L\mu+g_{RS(q)}\bar{e}P_R\mu)\bar{q}q\nonumber \\
	(g_{LV(q)}\bar{e}\gamma_\alpha P_L\mu+g_{RV(q)}\bar{e}\gamma_\alpha P_R\mu)\bar{q}\gamma^\alpha q\bigg]+\rm h.c \label{eq:conversionLagr}
\end{align*}
where $F^{\alpha\beta}$ is the photon field tensor. Contributions that depend on the spin-state of the nucleus arise from contact interactions involving axial-vector, pseudo-scalar and tensor quark currents. The spin-independent rate is dominant because it is enhanced by the coherent sum over all nucleons. The effective Lagrangian at the quark-level can be matched onto interactions involving nucleons via the following matrix elements
\begin{equation*}
	\ev{\bar{q}\Gamma_K q}{N}=G^{(p,q)}_K\bar{p} \Gamma_K p\qquad \ev{\bar{q}\Gamma_K q}{N}=G^{(n,q)}_K\bar{n} \Gamma_K n 
\end{equation*}
where $\Gamma_S=1, \Gamma_V=\gamma^\alpha$ and $p,n$ label protons and neutrons, respectively. The vector coefficients are obtained by the quark content of the nucleon $G^{(p,u)}_V=G^{(n,d)}_V=2$, $G^{(p,d)}_V=G^{(n,u)}_V=1$, $G^{p(n),s}_V=0$ while the scalar charges $G_S$ are extrapolated with dispersive relations and lattice results \cite{MatchingNucleon, MatchingNucleon2}. The wave function for the muon-bound state is calculated by solving the Dirac equation in the presence of the electric field of the nucleus and is averaged over the proton and neutron densities. Defining $\tilde{g}_{XK}^{(p)}=\sum_q G^{(p, q)}_{K}g_{XK(q)}$ and $\tilde{g}_{XK}^{(n)}=\sum_q G^{(n, q)}_{K}g_{XK(q)}$, the conversion rate reads
\begin{equation*}
	\Gamma_{\mu N\to e N}=2G^2_F\abs{A_L D + \tilde{g}_{LS}^{(p)}S^{(p)}+\tilde{g}_{LS}^{(n)}S^{(n)}+ \tilde{g}_{LV}^{(p)}V^{(p)}+\tilde{g}_{LV}^{(n)}V^{(n)}}^2 + L\leftrightarrow R \label{eq:conversionrate}
\end{equation*}
where $D,S,V$ are the overlap integrals in Equations~(19)--(23) of \cite{MutoeKitano}, that involve proton/neutron densities and muon/electron wave functions. All the overlap integral scale with the atomic number $\sim Z$, giving the anticipated coherence factor $Z^2$ that enhances the spin-independent $\mu\to e$ conversion rate over the muon capture rate.  
\externalbibliography{yes}

\reftitle{References}

\begin{adjustwidth}{-\extralength}{0cm}

\end{adjustwidth}

\end{document}